\documentclass[prd,twocolumn, nofootinbib, longbibliography,
]{revtex4-1}
\usepackage{hyperref}
\usepackage{tabularx}
\usepackage{amsmath}
\usepackage{float}
\usepackage{datetime}
\usepackage{textpos}
\usepackage{booktabs}
\usepackage{graphicx}
\usepackage{mathrsfs}

\newcommand{\apjs}{ApJS}

\newcommand{\aap}{A{\&}A}

\newcommand{\mnras}{MNRAS}

\begin{document}
\title{
The Contribution of Cosmic Rays Interacting With Molecular Clouds to the Galactic Center Gamma-Ray Excess}
\author{Oscar Macias}
\author{Chris Gordon}

\affiliation{Department of Physics and Astronomy, Rutherford Building, University of Canterbury, Private Bag 4800, Christchurch 8140, New Zealand}

\begin{abstract}
The Fermi-LAT data  appear to have an excess of gamma rays from the inner 150 pc of the Galactic Center. The main explanations proposed for this are: an unresolved population of millisecond pulsars (MSPs), dark matter (DM) annihilation,  and  nonthermal bremsstrahlung produced
by a population of electrons interacting with neutral gas in molecular clouds. 
The first two options have spatial templates well fitted by the square of a generalized Navarro-Frenk-White (NFW)  profile with inner slope $\gamma=1.2$.
We model the third option with a 20-cm continuum emission Galactic Ridge template. A template based on the HESS residuals is shown to give similar results.
The gamma-ray excess is found to be best fit by a combination of the generalized NFW squared template and a Galactic Ridge template. We also find the spectra of each template is not significantly affected in the combined fit and is consistent with previous single template fits. That is the generalized NFW  squared spectrum can be fit by either of order 1000 unresolved MSPs or DM with mass around 30 GeV,   a thermal cross section, and mainly annihilating to $b \bar{b}$ quarks. While the Galactic Ridge continues to have a spectrum consistent with a population of nonthermal electrons whose spectrum also provides a good fit to synchrotron emission measurements. We also show that the current DM fit may be hard to test, even with 10 years of Fermi-LAT data, especially if there is a mixture of DM and MSPs contributing to the signal, in which case the implied DM cross section will be suppressed.

\end{abstract}

\maketitle

\section{Introduction}
\label{sec:introduction}

Gamma rays constitute an excellent search channel for a signature of pair annihilation of dark matter (DM), since they can propagate almost without absorption from the source to the observer. Amongst all possible target regions in the gamma-ray sky, the Galactic Center is expected to be the brightest DM emitting source as it is relatively close by and has a high density of DM. However, this region is populated by a variety of astrophysical gamma-ray sources that make it hard to unambiguously identify a DM signal~\citep{cirelli2,Funk}. 

Several independent groups have reported evidence of extended excess 
 gamma-ray emission  above the diffuse galactic background (DGB) from the central $1^{\circ}-2^{\circ}$ around the Galactic Center~\cite{Goodenough:2009gk,Hooper:2010mq,Boyarsky:2010dr,hooperlinden2011,
 hooperkelsoqueiroz2012,
 AK,AKerratum,GordonMacias2013}. These investigation were based on Fermi Large Area Telescope (LAT) data. Although the Fermi-LAT Collaboration have not yet published a full Galactic Center analysis, in a preliminary study with one year of data, the Fermi team has reported an excess in observed counts peaking at energies of $\sim 2-5\;\rm{GeV}$~\cite{Vitale:2009hr,2011NIMPA.630..147V}. Given that there is a reasonable consensus on the reality of these Galactic Center excess gamma rays (GCEG), various alternative explanations for its origin have been posited:

(i) DM particles with masses of about $10- 100\ \mathrm{GeV}$ annihilating into $b\bar{b}$ and $\tau^+\tau^-$ final states or a combination of both~\cite{Goodenough:2009gk,Hooper:2010mq,hooperlinden2011,hooperkelsoqueiroz2012,AK,AKerratum}. Importantly, it was argued  in Ref.~\citep{GordonMacias2013} that the signal has a relatively soft spectral shape, which makes it difficult to fit the GCEG data with a dark matter model annihilating mainly to leptons. The spatial profile of the DM was found to be well fit \citep{GordonMacias2013} by a generalized NFW   profile \citep{ioccopatobertone2011} with inner slope $\gamma=1.2$. As
the DM signal is proportional to $\rho^2$, the spatial profile used will be the 
square  of a generalized NFW   profile with inner slope $\gamma=1.2$. We will denote this spatial profile as  (NFW$_{1.2}^2$).

(ii) A superposition of $\sim 10^3$ millisecond pulsars (MSPs) within a radius of $r\lesssim 150$ pc of the Galactic Center whose number density follow a NFW$_{1.2}^2$ profile~\cite{Abazajian:2010zy,AK,AKerratum,wharton2012,GordonMacias2013,GordonMacias2013erratum,Miribal2013}. 
However,  Refs.~\citep{hooperslatyer2013, Huang:2013} have claimed that there is evidence of a gamma-ray excess at $2$~kpc~$\leq r\leq 3$~ kpc that  is consistent with DM annihilation 
but is too extended to be  explained by a concentrated population of MSPs  given the number of MSPs that have been resolved by Fermi-LAT \citep{Hooper:Pulsars}.

(iii) Another possibility is that the signal is being produced by cosmic rays interacting with gas in the Galactic Center \cite{Goodenough:2009gk,hooperlinden2011,Linden:2012iv,yusef-zadehhewittwardle2013,AK,AKerratum}. This alternative solution can be divided in two different scenarios, the hadronic and nonthermal bremsstrahlung. The first one consists of $\pi^0$-decays resulting from the emission of high energy protons  and their subsequent collision with   gas in the Galactic Center.
 In Ref.~\cite{Linden:2012iv} it was found  that a model based on hadronic emission from Sgr A* would be determined predominately by the gas distribution and would appear point-like to the Femi-LAT gamma-ray detector. Therefore, that model would not be suitable for explaining the  extended nature of the GCEG. 
 
 In the second scenario, the nonthermal bremsstrahlung emission model, 
 a case which results in extended emission has been proposed in Ref.~\citep{yusef-zadehhewittwardle2013}.
 Based on multi-wavelength observational data obtained with the Green Bank Telescope (GBT) \cite{law2008}, Susaku, X-ray Multi-Mirror Mission (XMM)-Newton, Chandra, Fermi-LAT and High Energy Stereoscopic System (HESS) it was argued~\citep{yusef-zadehhewittwardle2013} that the $\sim$GeV GCEG is nonthermal, diffuse and is probably generated by a population of synchrotron emitting electrons interacting with gas in molecular clouds. 

In this study we focus on the spatial and spectral morphology of the gamma-ray Galactic Ridge (hereafter ``Galactic Ridge'') region, and confirm that an extended source associated with the Galactic Ridge can improve the GCEG fit. But, we find that      adding a Galactic Ridge does not remove the need for also adding a spherically symmetric extended source whose radial profile follows a NFW$_{1.2}^2$ profile. We show that the spectral parameters of the NFW$_{1.2}^2$ template are not significantly affected by inclusion of a Galactic Ridge.

\section{Data Reduction}
\label{sec:data}

The Fermi-LAT data selection is the same as described in~\citep{GordonMacias2013}. In summary, we analysed Pass-7\footnote{Preliminary checks have shown our results are not significantly changed if we instead use Pass-7 reprocessed  data.} data taken within a squared region of $7^{\circ}\times7^{\circ}$ centred on Sgr A$^{\star}$ in the first 45 months of observations over the period August 4, 2008$-$June 6, 2012. We used the standard data cuts and kept only the \texttt{SOURCE} class events which have a high probability of being photons of astrophysical origin. We also selected events between 200 MeV$-$100 GeV without making any distinction between \textit{Front} and \textit{Back} events. 

The spectra were obtained by maximizing the likelihood of source models using the binned \textit{pyLikelihood} library in the Fermi Science Tools~\citep{fermitools}. We followed the same fitting procedure adopted in Ref.~\citep{AK,AKerratum} which has been recommended to be more suitable for crowded regions like the Galactic Center. Unless otherwise stated, the models included all sources suggested in the 2FGL~\citep{2FGL} catalog plus the LAT standard DGB and extragalactic background models \texttt{gal$_{-}$2yearp7v6$_{-}$v0.fits}  and  \texttt{iso$_{-}$p7v6source.txt} respectively.

\section{Models for the Extended Source at the Galactic Center }
\label{sec:Models}

The HESS telescope has revealed a point-source coinciding with the dynamical center of the Milky Way Galaxy as well as diffuse emission that is spatially correlated with the molecular clouds in the Galactic Ridge~\cite{Aharonian:2006au}. In Ref.~\citep{yusef-zadehhewittwardle2013} it was argued that  bremsstrahlung from nonthermal electrons in Galactic Center molecular clouds can explain the GCEG  measured at TeV scales by HESS and at GeV scales by  Fermi-LAT. The non-thermal electrons in the molecular clouds are propoÁsed to mainly come from supernova remnants and nonthermal radio filaments (see \citep{law2008,yusef-zadehhewittwardle2013} and references therein).
A proposed population of nonthermal electrons is constrained, by both radio and gamma-ray data, to need a broken power law spectrum where the break is attributed to rapid cooling of electrons at high energies \citep{yusef-zadehhewittwardle2013}.
By comparing the frequency of the break in the radio data and the energy of the break in the gamma-ray data, the magnetic field
value can be constrained, see Sec.~\ref{sec:Results}. 

The TeV nonthermal electrons, proposed to explain for the HESS Galactic Ridge, are assumed to be a separate younger population of nonthermal electrons in the Galactic Center molecular clouds. This extra population is assumed  to have not had time to cool and so is modeled with a power law distribution \citep{yusef-zadehhewittwardle2013}.

To study the evidence for a new component of extended GeV emission in the Fermi-LAT data, the authors in Ref.~\citep{yusef-zadehhewittwardle2013} tried spatial templates  obtained from X-ray, 20-cm continuum emission radio data, and the HESS residuals. For a spectral model they initially employed a broken power law of the form:
\begin{equation}
  \frac{dN}{dE}=N_0 \times 
  \begin{cases}
    \left(\frac{E}{E_b}\right)^{-\Gamma_1} &\text{if $E<E_b$} \\
\left(\frac{E}{E_b}\right)^{-\Gamma_2} &\text{otherwise.}
  \end{cases}
   \label{eq:BrokenPl}
\end{equation} They found that the 20-cm radio and HESS residual templates had similar high test statistic (TS) values.

\begin{figure*}[p!]Á
\begin{center}
\begin{tabular}{c}
\begin{minipage}{0.9\linewidth}
\hspace*{-0.6cm}\includegraphics[width=\linewidth]{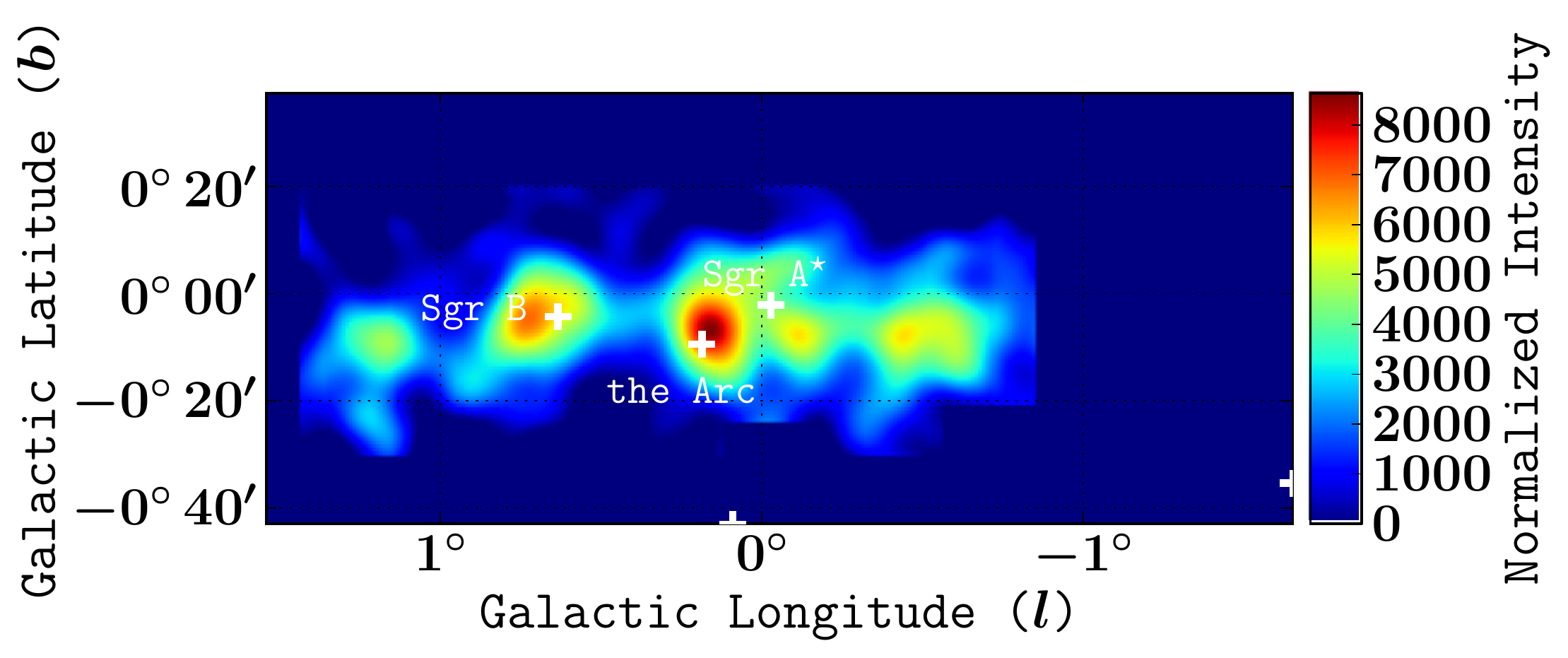}
\end{minipage}
 \\
 \begin{minipage}{\linewidth}
 \includegraphics[width=\linewidth]{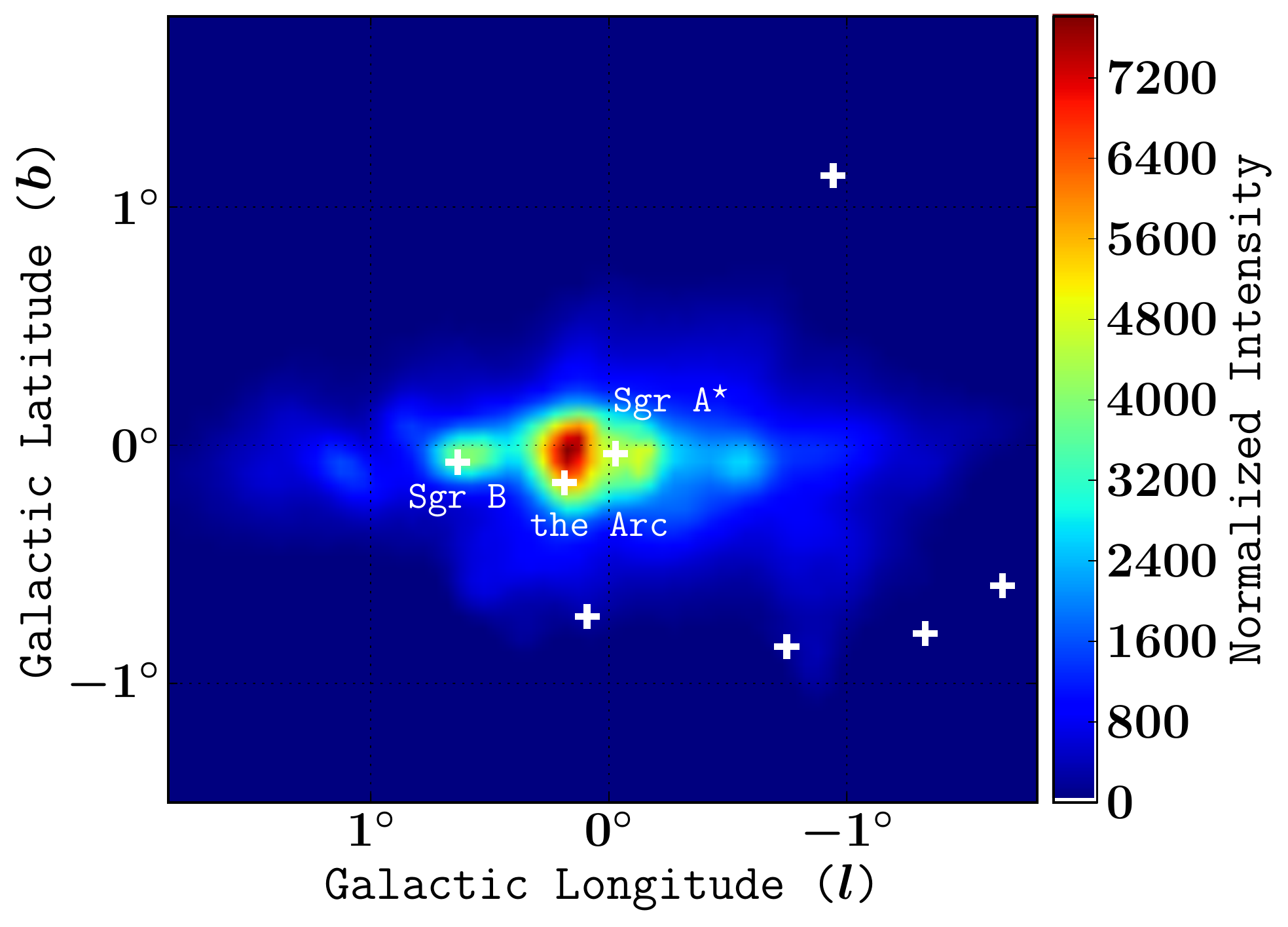} 
 \end{minipage}\\ 
\end{tabular}
\caption{ \label{fig:20-cmMap} \label{fig:HESSMap}\textit{Top:} Gamma-ray image of the Galactic Center as observed by the HESS telescope ($E>380$ GeV) after subtracting the dominant point sources~\citep{Aharonian:2006au}.
 To include this template map in the likelihood function within the Fermi Science Tools package, we background subtracted, thresholded and normalized the data provided in~\citep{Aharonian:2006au}.  \textit{Bottom:}
 Background-subtracted, thresholded, and normalized image from 20-cm continuum emission GBT data. This template was 
the same one as used in  Ref.~\citep{yusef-zadehhewittwardle2013} and we refer the reader to that article for details\footnote{We thank Farhad Yusef-Zadeh for providing us with the 20-cm template.}. This spatial template is named the ``20-cm template'' in the rest of this work.   The crosses overlaid on the image represent the position of the 2FGL catalog point sources. 
 }   
\end{center}
\end{figure*}

For illustration, we show in Fig.~\ref{fig:HESSMap}  the HESS residual and 20-cm spatial templates.
The 20-cm template was based on  GBT continuum emission data  which measures nonthermal and thermal plasma distributions \citep{law2008,yusef-zadehhewittwardle2013}.  Note,  this is  distinct from the 21-cm line temperatures used by the Fermi team in constructing the DGB as that gives a measure of the column density \citep{DGB}.
 Both templates initially had a DC value, evaluated from a nearby region, subtracted. They have also  had  Sgr A  removed and they have been normalized so that their total area integrated flux is unitary.

To test whether the GCEG is better fitted by  a combination of a NFW$_{1.2}^2$ template and 
a Galactic Ridge template
we have done a broad band analysis within the Fermi Tools and also a bin-by-bin analysis for each of the extended sources under scrutiny.

\section{Systematic errors and parameter constraints}
\label{sec:Systematics}
 
The DGB  accounts for a large proportion of the photons detected by the LAT instrument. For regions near the Galactic Center this component can be several orders of magnitude brighter than any other source. In particular, the dominant systematic error at energies $\sim$1 GeV emerges from the uncertainties in the DGB model. These systematics were studied in a previous analysis~\citep{GordonMacias2013}.

Since this work involves the analysis of an extra extended source (see section~\ref{sec:Models}) not considered in~\citep{GordonMacias2013}, we have reassessed the systematic errors in the DGB by following the same approach explained in~\citep{GordonMacias2013}. There is consistency between the present and previous analysis~\citep{GordonMacias2013}, we found that the overall systematic flux error is energy and spatial dependent: systematic errors due to uncertainties of the spectral distribution amounts to an average of about $2\%$ at $\sim$1 GeV, and the dominant fraction for the systematics arises from the spatial part, we obtained on average about 23\% for energy bins $\leq$ 10 GeV and $18\%$ in the 10$-$100 GeV energy band. 
The total systematic error is evaluated by summing in quadrature the spatial, spectral, and effective area \citep{2FGL,GordonMacias2013} systematic errors. 

Our parameter constraints method is the same as used in Ref.~\citep{GordonMacias2013}. In summary we use the Fermi Tools to construct a spectrum  of the source of interest \citep{2FGL}.
As in
Refs. \citep{2FGL,GordonMacias2013}, we allow the amplitude of all sources, in the region of interest, to vary when fitting a band.
 We then add, in quadrature, the systematic errors (evaluated as described above)  to the statistical errors of the spectral bands. The spectrum likelihood is then approximated as a multivariate Gaussian and a profile likelihood approach is used to construct confidence intervals.

In plotting the spectra we display both the systematic and statistical errors. For bands which have a test statistic (TS) 
\citep{2FGL,GordonMacias2013}
less than 10, or whose total error is more than the half of the best fit band value, we plot the 95\% upper limit. We do not plot or use bands in our parameter constraints which have TS$<1$. Unless otherwise stated, best fit parameter values are quoted with 68\% confidence intervals.

To cross-check the systematical errors explained above, we have also estimated the systematic uncertainties in the DGB following the interesting analysis technique utilized in~\citep{ackermannajelloatwood2012}. We constructed eight different diffuse emission models using GALPROP~\citep{GALPROP, webrun}, and each of these templates were included in the likelihood fit of the sources of interest as an alternative to the standard DGB recommended in the 2FGL catalog~\citep{2FGL}. 

\begin{figure*}[ht!]
\begin{center}

\begin{tabular}{cc}
\centering
\includegraphics[width=0.5\linewidth]{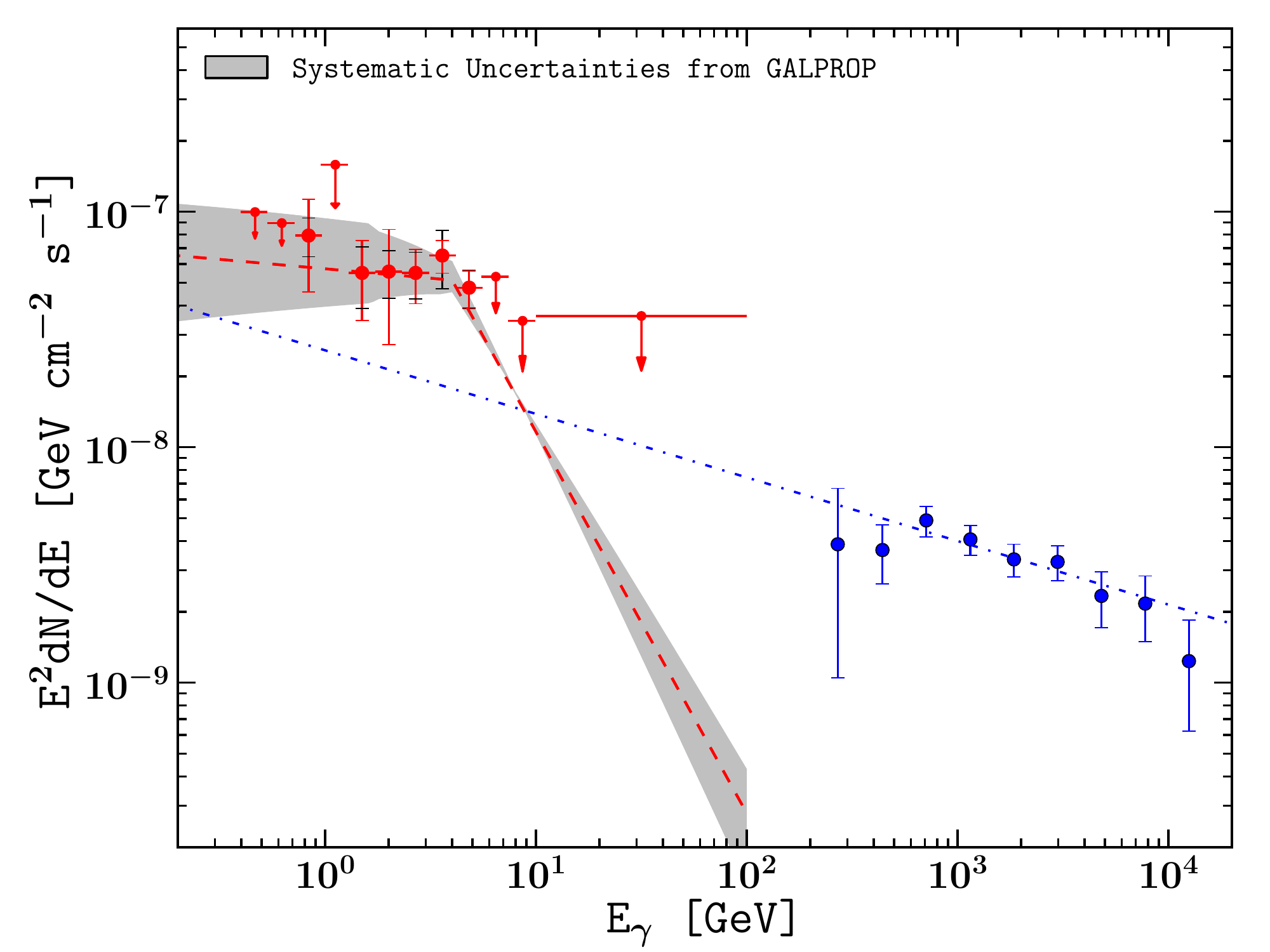} & \includegraphics[width=0.5\linewidth]{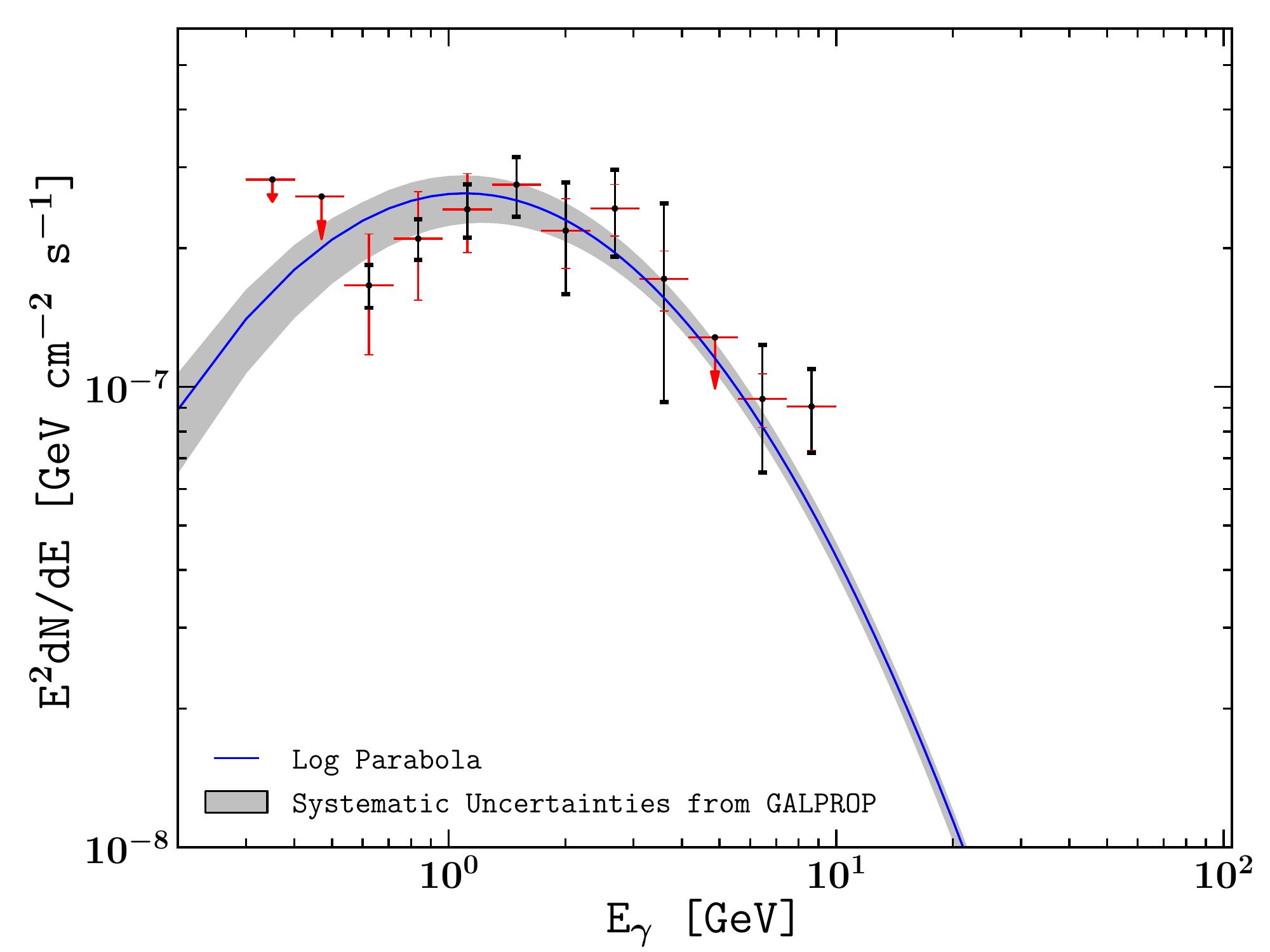}  
\end{tabular}
\includegraphics[width=0.5\linewidth]{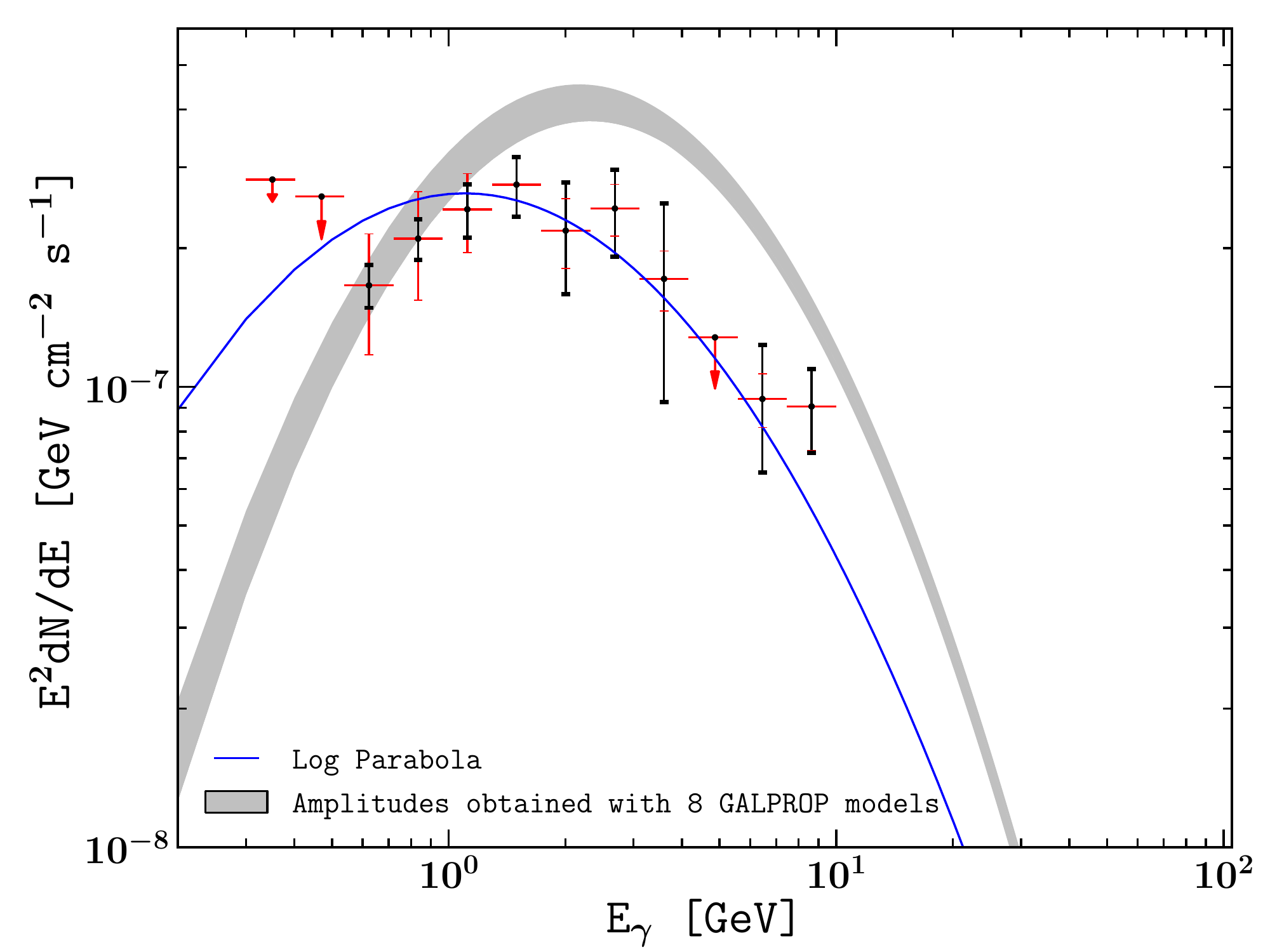}

\caption{ \label{fig:galpropuncertainties}\textit{Top Left:} Galactic Ridge spectrum generated from model 2 in Table~\ref{tab:LogLikelihoods}. 
 The red dashed line shows the best fit broken power-law as obtained from a Fermi Tools broad band fit. The grey area is an estimate of the systematic uncertainties as calculated with 8 different GALPROP models of the DGB. Black and red error bars are the LAT ($1\sigma$) statistical and systematic errors. A red arrow indicates a 95\% upper limit.  The blue points correspond to data taken by HESS \cite{Aharonian:2006au} and the blue dotted line is the best fit power law to them. \textit{Top Right:} spectrum for NFW$_{1.2}^2$ spatial profile
 generated with model 1 in Table~\ref{tab:LogLikelihoods}.
  The blue solid line is the best fit Fermi tools log parabola spectrum. The grey area shows the systematic uncertainties as computed with 8 different GALPROP models of the DGB.  The spectra and error bars are listed in Table~\ref{tab:DMdata20-cmMap} and \ref{tab:Ridgedata20-cmMapNOArcandSgrB}. \textit{Bottom Center:} Same as top right, except the  GALPROP based model results are shown rather than  just the relative errors obtained from them.}
\end{center}
\end{figure*}

The set of alternative DGB models taken into account in this analysis consider a range of possible values for the input parameters that were found to exhibit the largest sensitivities~\citep{ackermannajelloatwood2012}. The parameters varied in the models are the cosmic ray propagation halo heights (4 kpc or 10 kpc), cosmic ray source distribution (supernova remnants or pulsars) and the atomic hydrogen spin temperature ($150$\,K or optically thin). An $E(B-V)$ magnitude cut of 5 mag was also chosen. The results obtained through this method are displayed as grey shaded areas in the spectra of Fig.~\ref{fig:galpropuncertainties}.

The  DGB provided with the Fermi Tools is generated by a weighted sum of gas column densities and an inverse Compton intensity map~\citep{DGB}. In the DGB generation, the weights depend on the energy band  and the gas template weights  also  depend on the ring radius concentric around the Galactic Center. The weights are fitted to all sky Fermi-LAT data. Due to the greater degree of freedom this method produces a better fit to the Fermi-LAT data than the GALPROP based approach described above. So in general the GALPROP simulations do not envelope the solution found using the standard DGB. Therefore we use the relative dispersion of the GALPROP simulations 
 in constructing the grey bands in the top panels of Fig~\ref{fig:galpropuncertainties}. In the bottom panel of Fig~\ref{fig:galpropuncertainties} we plot the band of solutions obtained when the GALPROP DGB's are used. In this panel, the NFW$_{1.2}^2$ template has a greater amplitude as the GALPROP estimate of the DGB is not as good a fit as the standard DGB provided with the Fermi Tools.

\section{Results}
\label{sec:Results}

 As seen from Fig.~\ref{fig:20-cmMap}, the Arc and Sgr B are bright sources in the Galactic Center. They are thought to be associated with cosmic rays interacting with molecular clouds \citep{yusef-zadehhewittwardle2013} and so in Table~\ref{tab:LogLikelihoods} we consider models with and without them being assumed to be  included in the Galactic Ridge template.
 
 The results listed in Table~\ref{tab:LogLikelihoods} show that
 the broad band analysis revealed significant detections of
both a Galactic Ridge and a NFW$_{1.2}^2$ extended source.

\begin{table*}[ht!]
\begin{ruledtabular}
\begin{tabular}{lrr}
\centering
Model & $2\log (\mathcal{L}/\mathcal{L}_{\rm base})$ & dof$-$dof$_{\rm base}$\\  \hline 
Base (2FGL$-$``the Arc''$-$Sgr B)&  0 & 0  \\ \hline
2FGL& 425 &   4+5=9  \\
2FGL$+$20-cm template & 638 &4+5+4=13\\ 
2FGL$+$NFW$_{1.2}^2$    & 1295  & 4+5+3=12 \\
2FGL$+$NFW$_{1.2}^2$ $+$ HESS residual template   & 1325  & 4+5+3+4=16\\
2FGL$+$NFW$_{1.2}^2$ $+$ 20-cm template  ({\bf model 1})  & 1330 & 4+5+3+4=16 \\ \hline
Base$+$NFW$_{1.2}^2$ $+$ HESS residual template    & 1164 & 3+4=7 \\
Base$+$NFW$_{1.2}^2$ $+$ 20-cm template   ({\bf model 2}) & 1170  & 3+4=7 \\
\end{tabular}
\end{ruledtabular}

\caption{\label{tab:LogLikelihoods} The likelihoods evaluated in compiling the above table are maximized with a broad band analysis using the Fermi Tools. Alternatives models of the Galactic Center in the 200 MeV$-$100 GeV energy range are listed.
Each point source in the model has degrees of freedom (dof) from its spectrum and two extra dof from its location. The spectra for the Galactic Ridge templates are modeled by a broken power law. While the spectra for the NFW$_{1.2}^2$ templates are modeled by a log parabola which has enough flexibility to mimic a good fitting DM or MSP spectra \citep{GordonMacias2013}.
 }
\end{table*}

\begin{figure*}[p!]
\begin{center}

\begin{tabular}{ccc}
\centering
\includegraphics[width=0.33\linewidth]{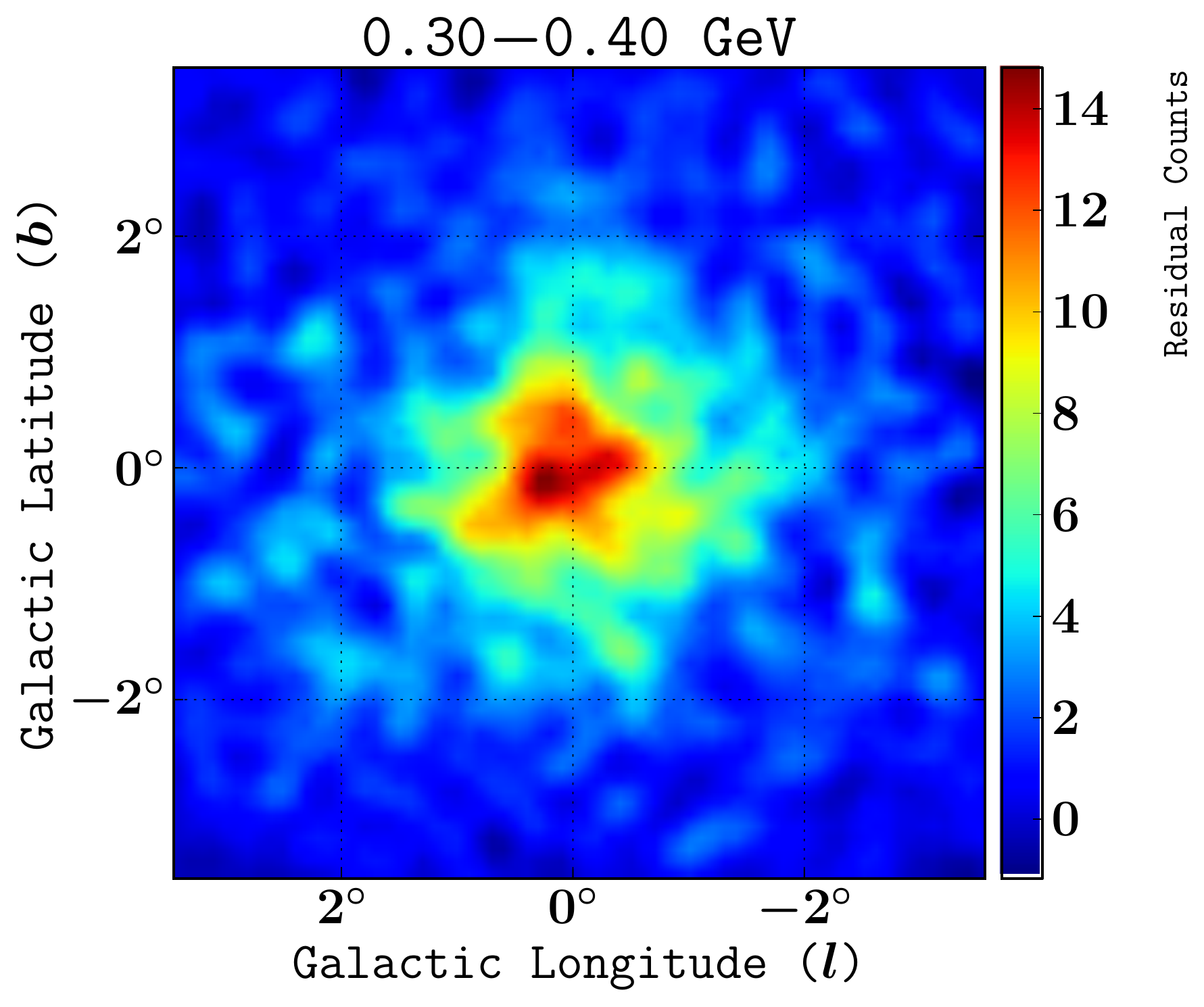} & \includegraphics[width=0.33\linewidth]{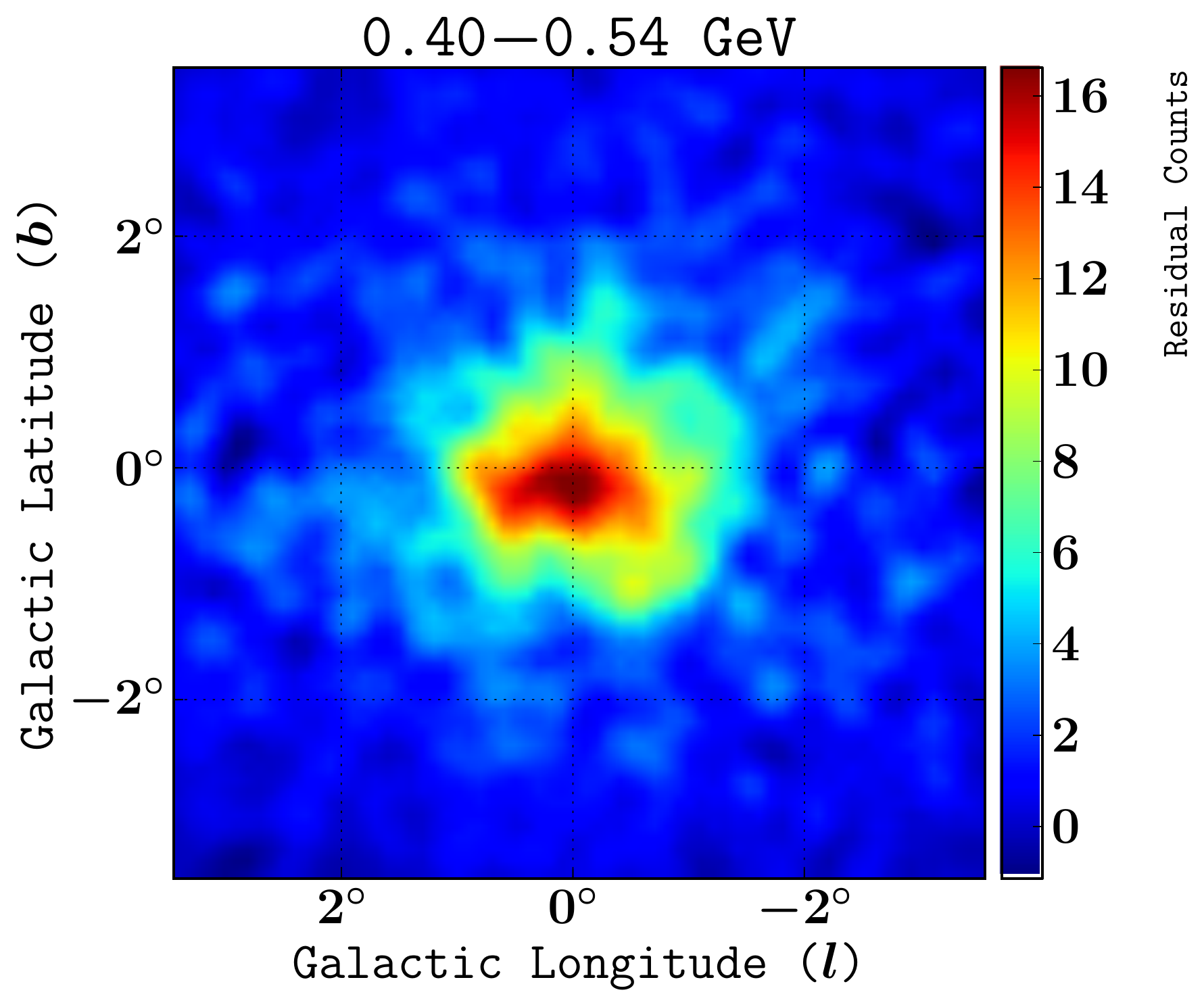} & \includegraphics[width=0.33\linewidth]{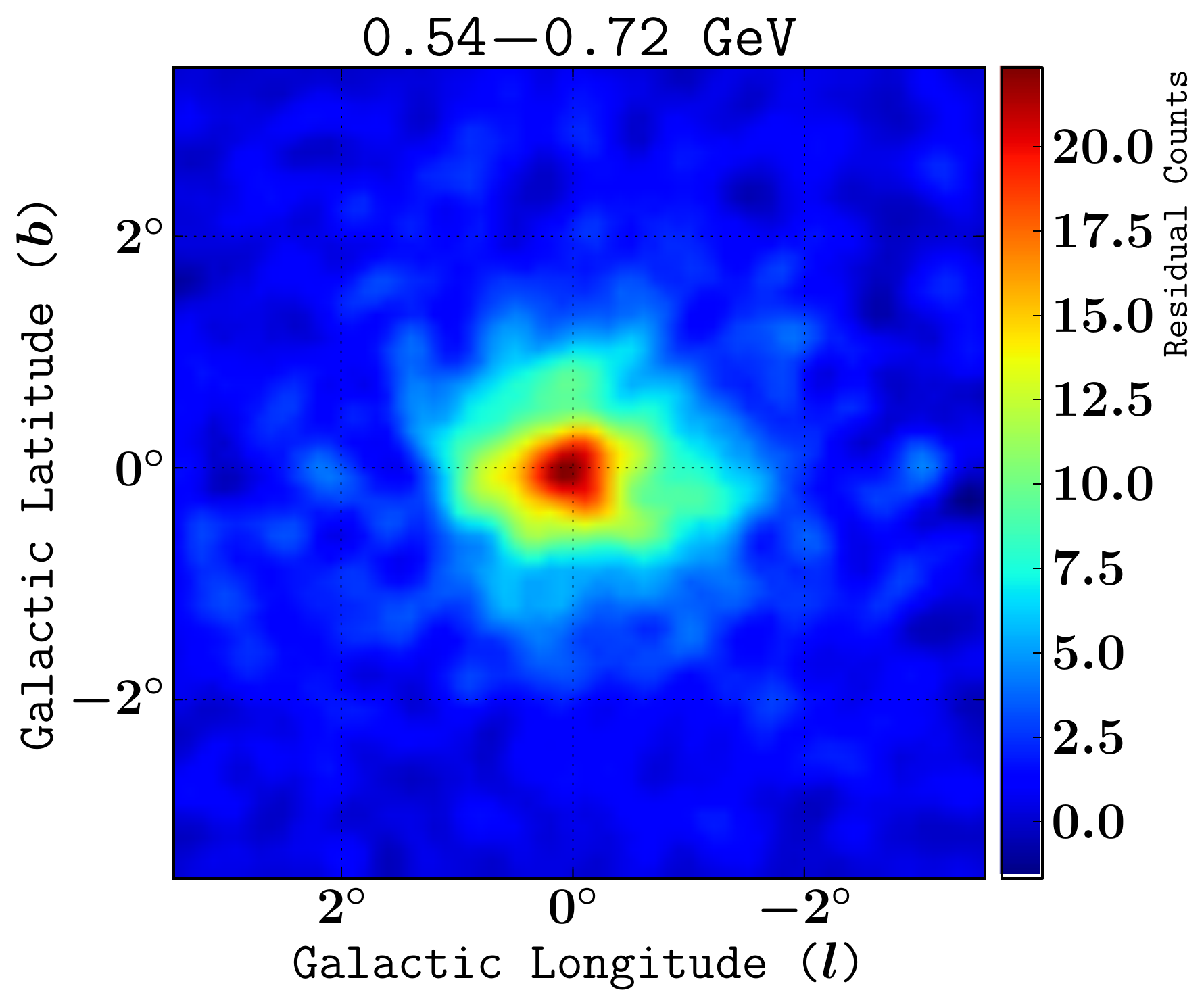}      

\end{tabular}

\begin{tabular}{ccc}
\centering
\includegraphics[width=0.33\linewidth]{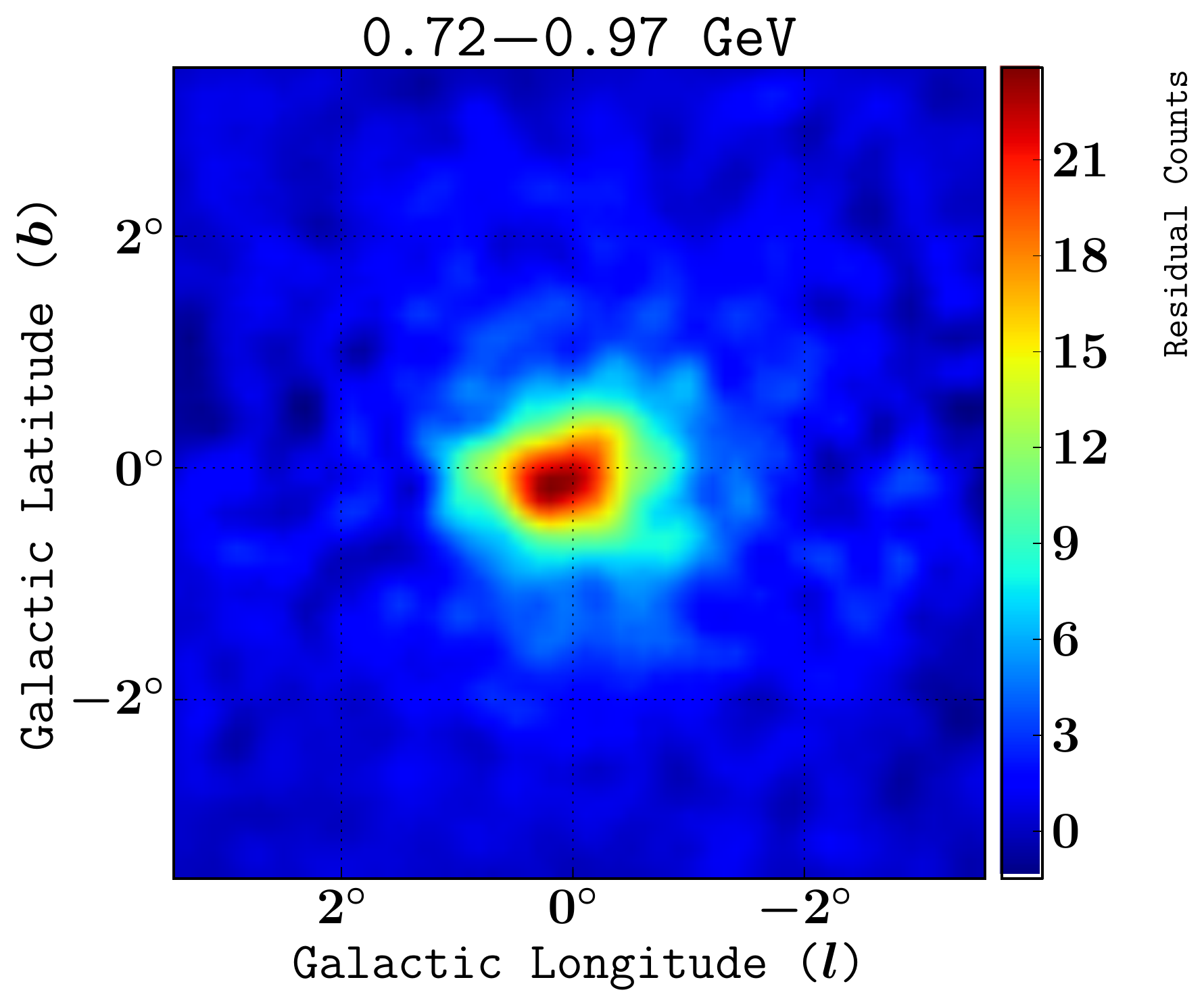} & \includegraphics[width=0.33\linewidth]{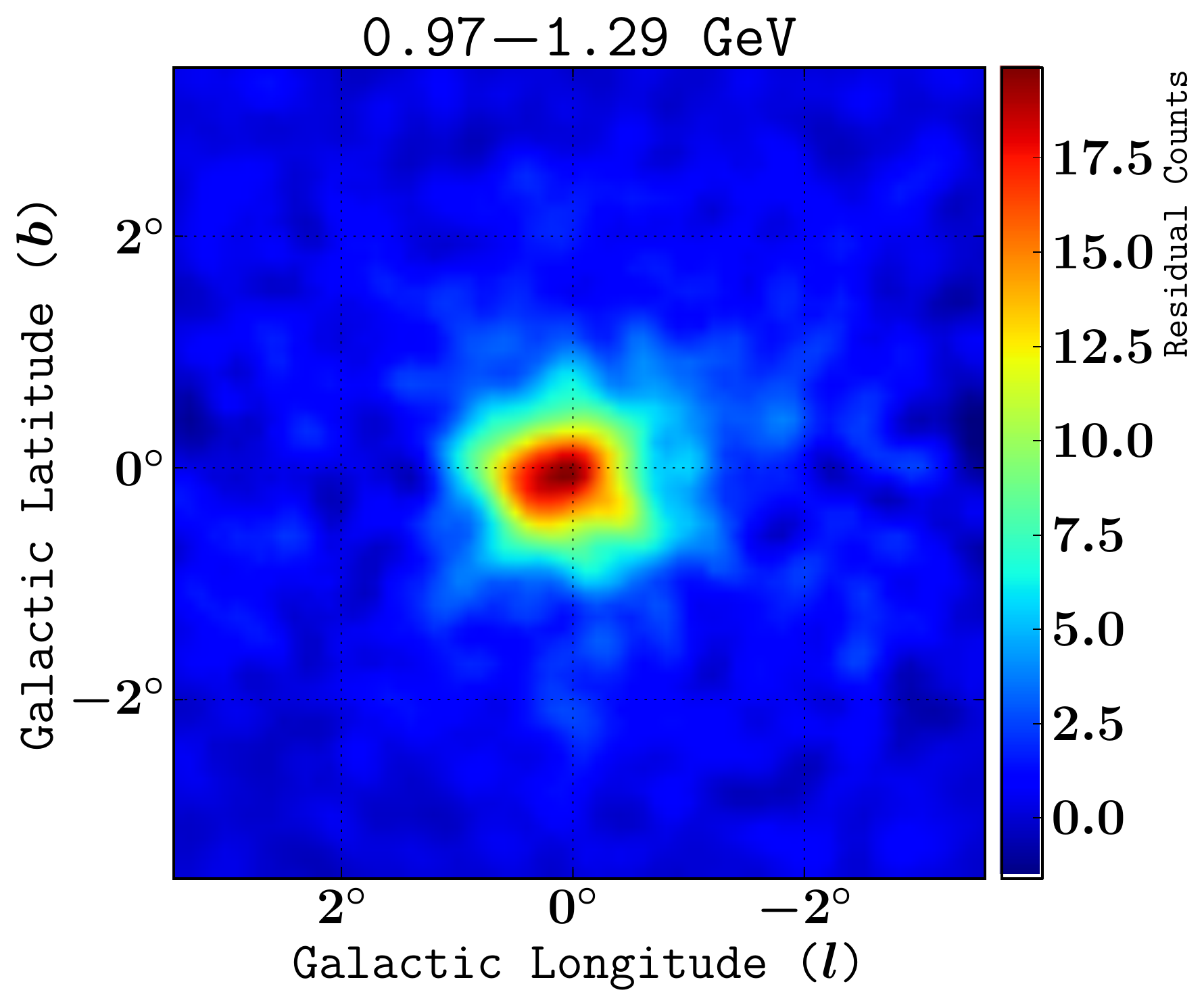} & \includegraphics[width=0.33\linewidth]{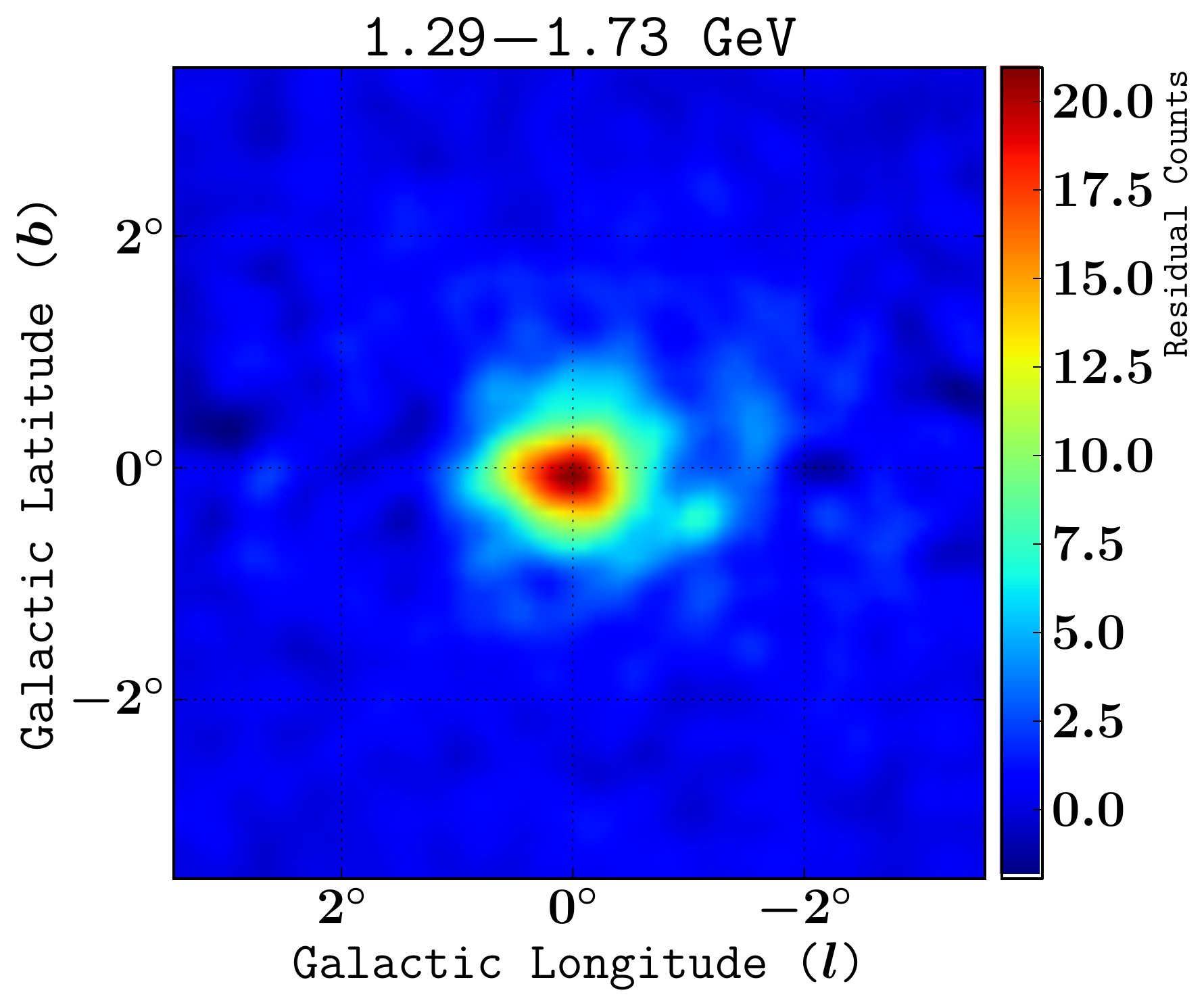}      

\end{tabular}

\begin{tabular}{ccc}
\centering
\includegraphics[width=0.33\linewidth]{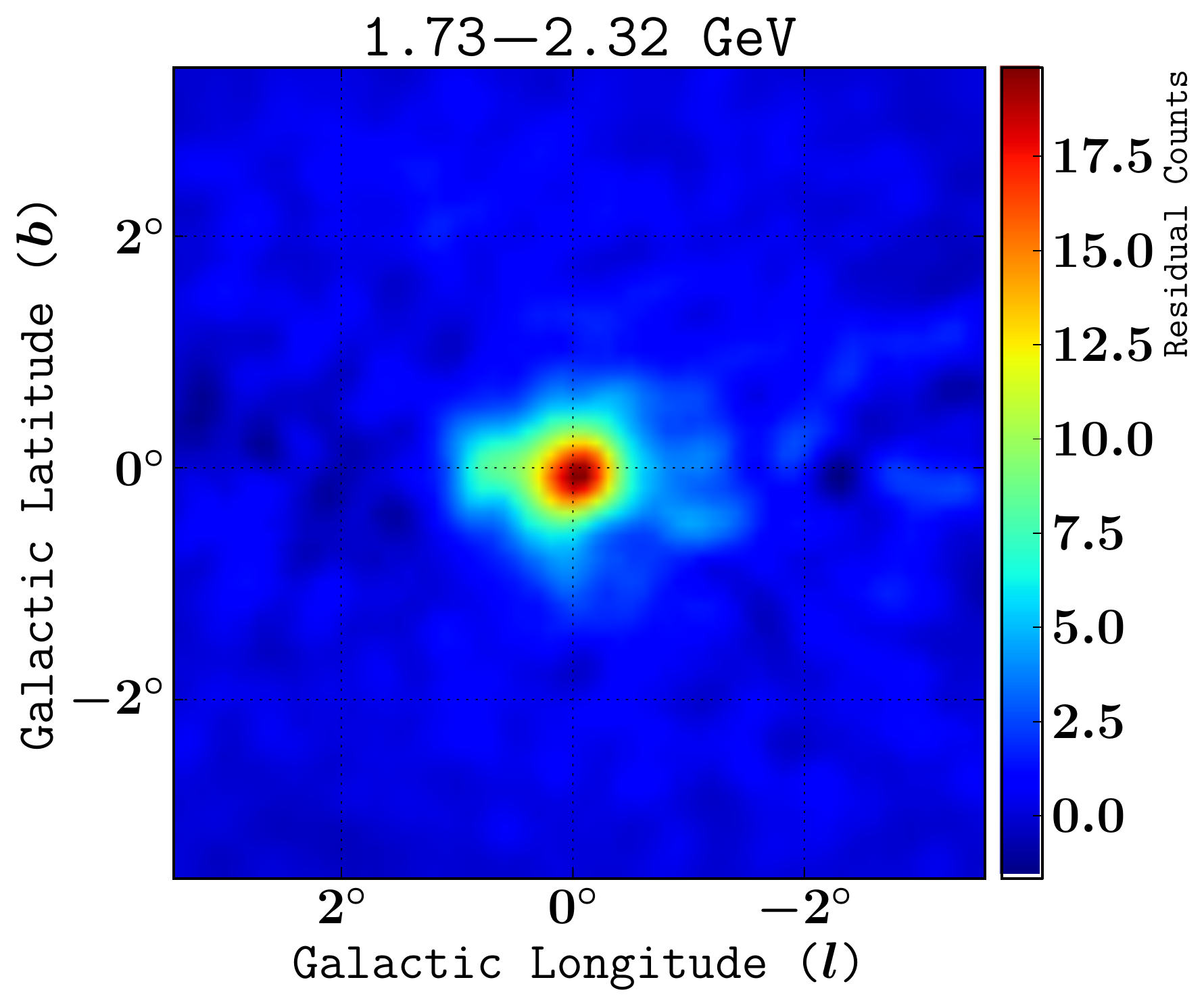} & \includegraphics[width=0.33\linewidth]{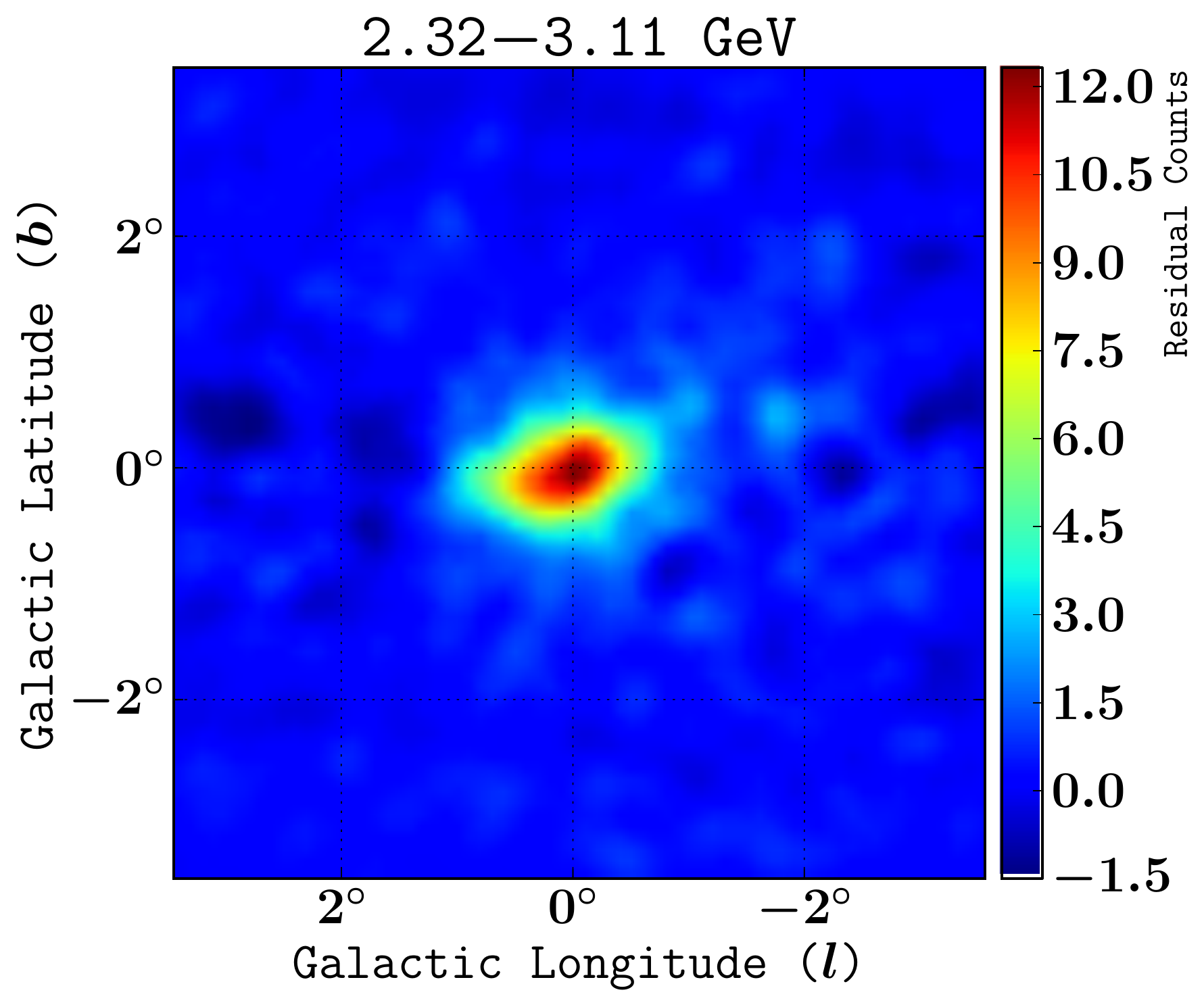} & \includegraphics[width=0.33\linewidth]{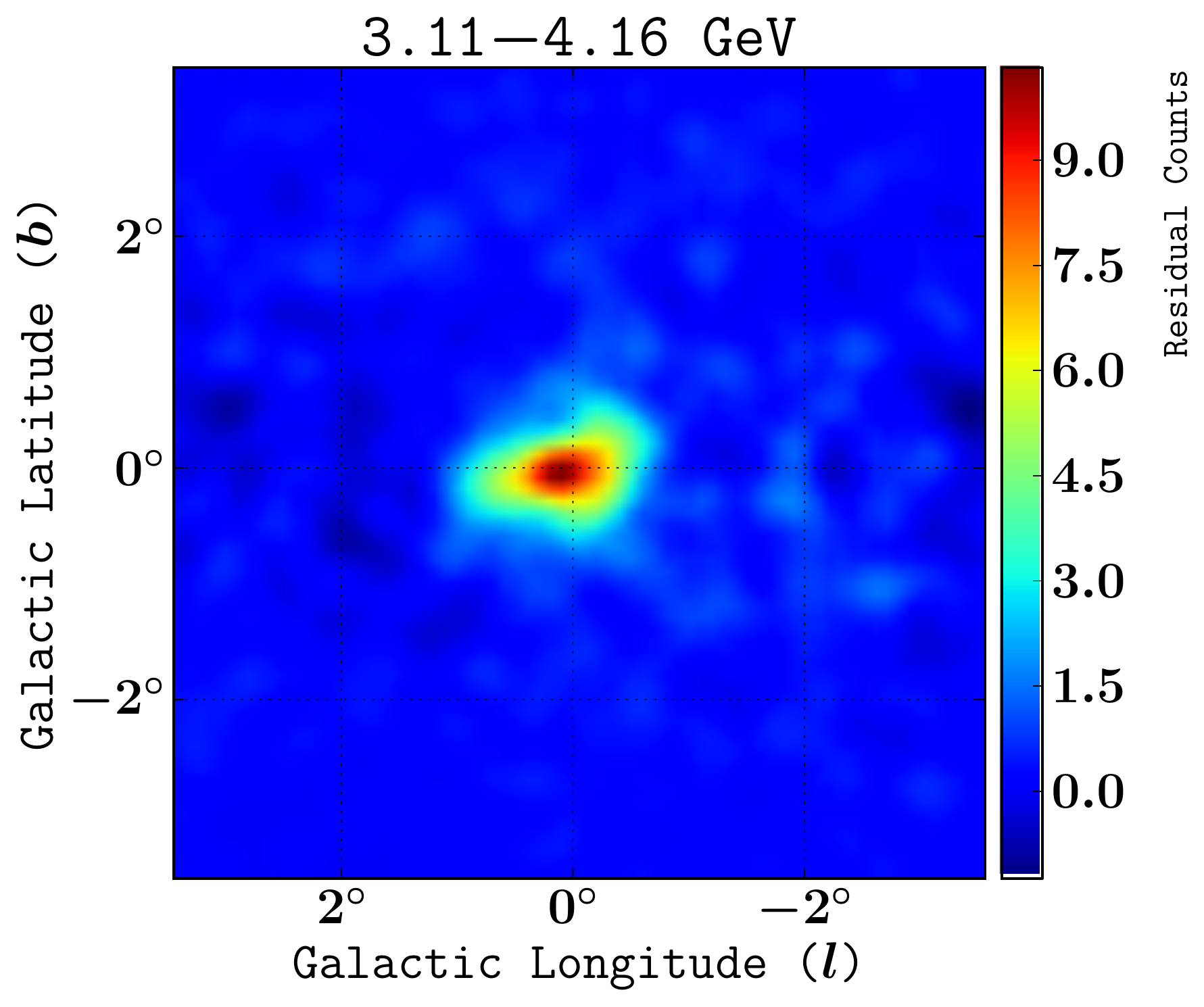}      

\end{tabular}

\begin{tabular}{lcc}

\includegraphics[width=0.33\linewidth]{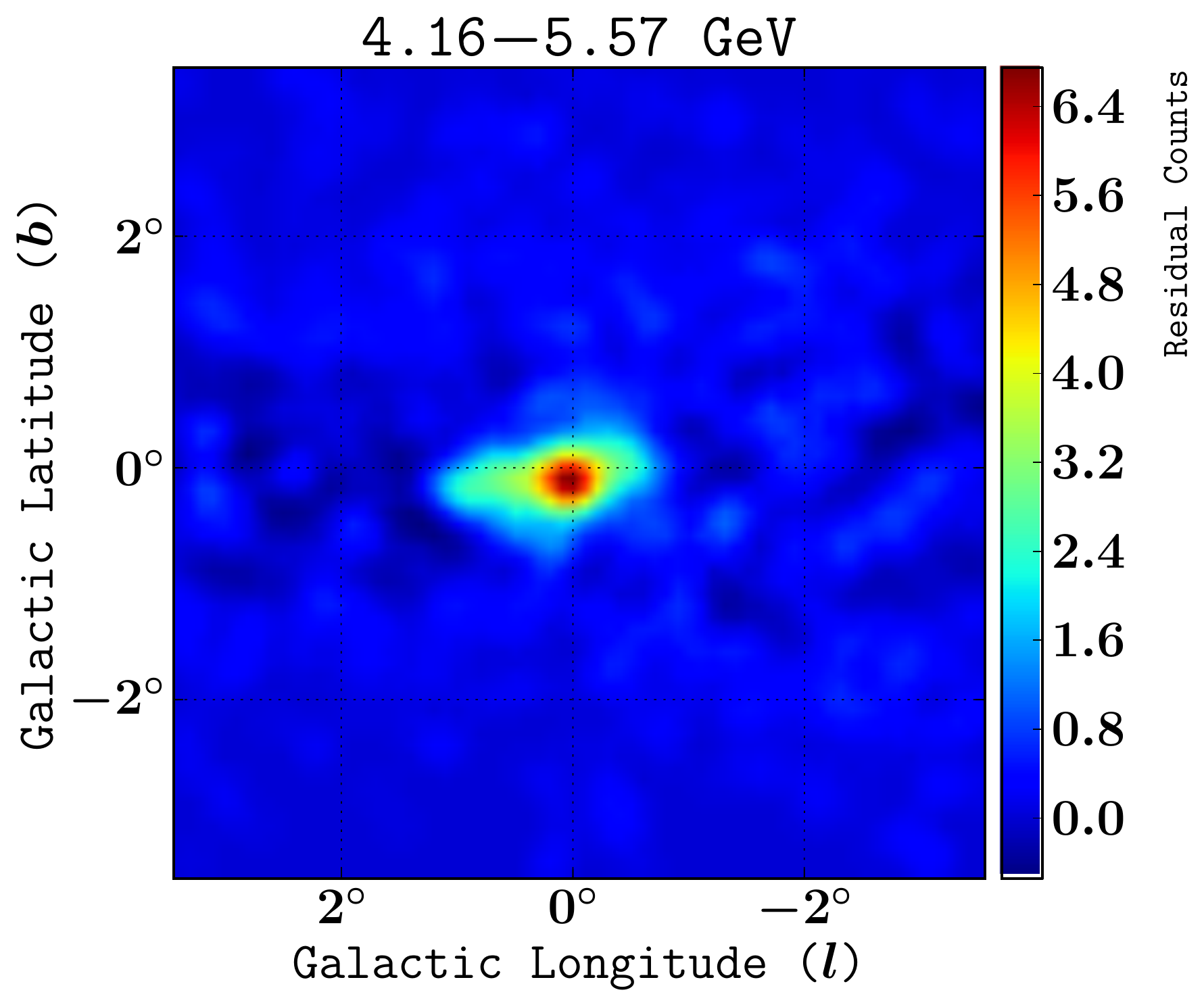} & \includegraphics[width=0.33\linewidth]{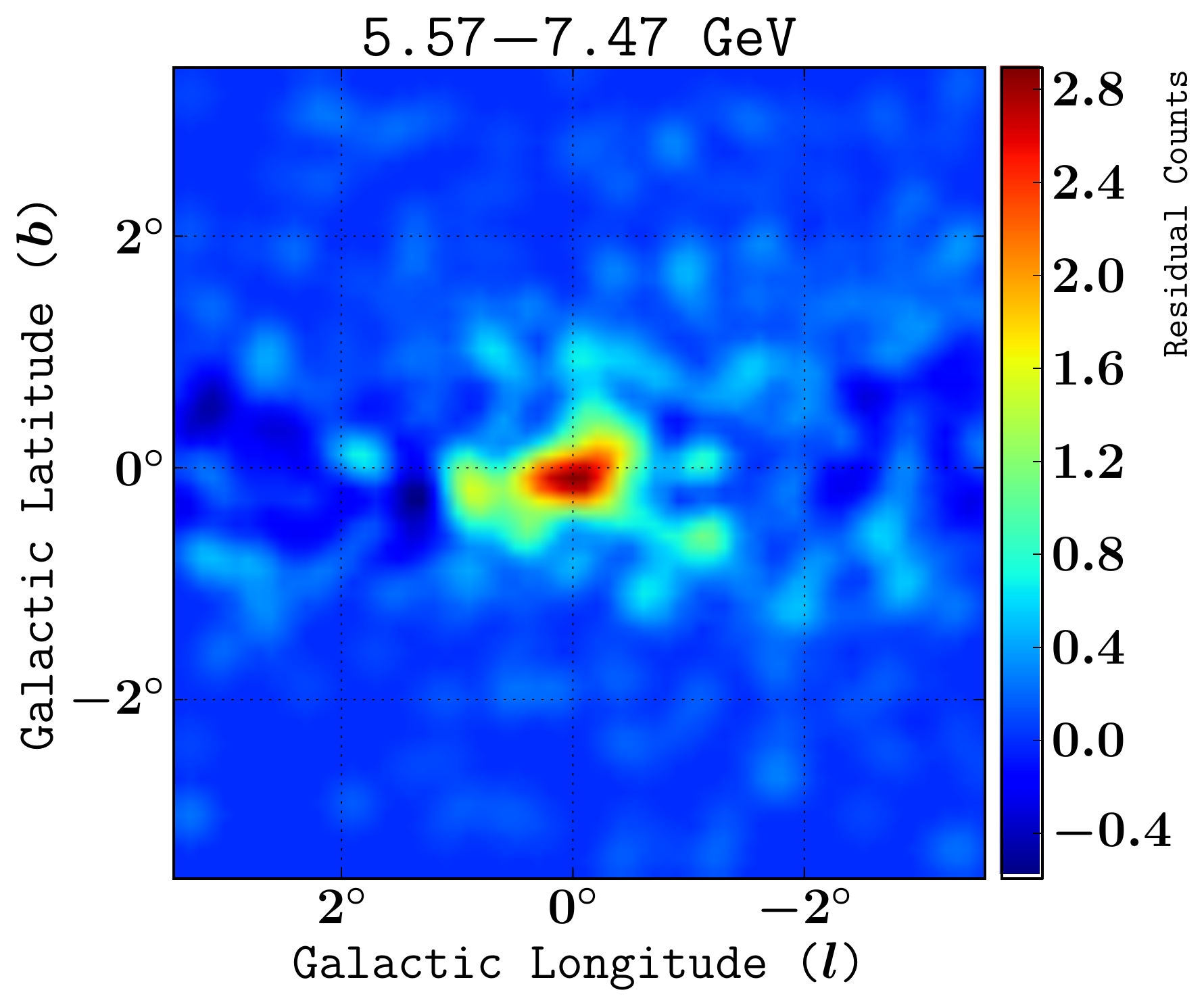}  & \includegraphics[width=0.33\linewidth]{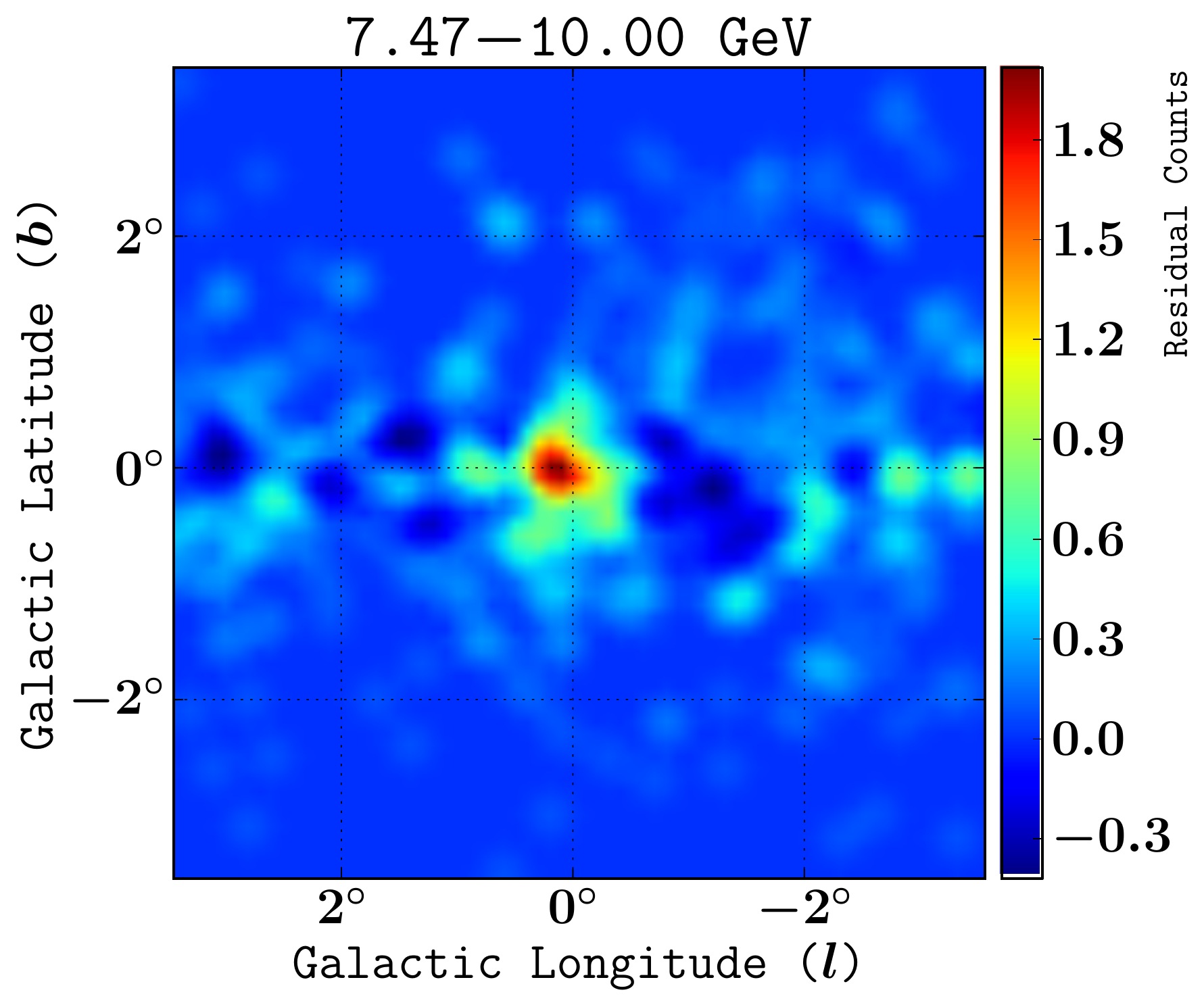}
\end{tabular}

\caption{ \label{fig:residuals} Shown are the residuals of model 2 in Table~\ref{tab:LogLikelihoods} where the model components of the NFW$_{1.2}^2$ and the 20-cm Galactic Ridge have not been subtracted from the data. The images have been smoothed with a 0.3$^\circ$ radius Gaussian filter.
}
\end{center}
\end{figure*}

The need for the Galactic Ridge can be seen in the residuals shown in Fig.~\ref{fig:residuals}. It is particularly noticeable in those bands which have a high TS (see Table~\ref{tab:Ridgedata20-cmMapNOArcandSgrB}).

Based on the GBT radio data,  Ref.~\citep{yusef-zadehhewittwardle2013} set the synchrotron flux  at 325 MHz to be $F_{325} =508$ Jy and  a synchrotron spectrum of electrons of the form $E^{-p}$ with $p=1.5$ below the break frequency $\nu_b=3.3$~GHz and $p=4.4$ above it. 
The GCEG spectrum can be used to constrain the break energy for the electron spectrum ($E_b$) via Eq.~\ref{eq:bremsstrahlungSpectra}. This can be converted to a constraint on the  magnetic field strength $B$  by using the measured radio frequency spectral break $\nu_b$ and the general relation between electron energy and characteristic  synchrotron radio frequency given in Eq.~\ref{eq:Enu}. The GBT uncertainties for the spectral slopes, $\nu_b$, and $F_{325}$ were not given in Ref.~\cite{yusef-zadehhewittwardle2013} and so our analysis just includes their point estimates.

\begin{figure*}
\begin{center}

\begin{tabular}{cc}
\centering
\includegraphics[width=0.47\linewidth]{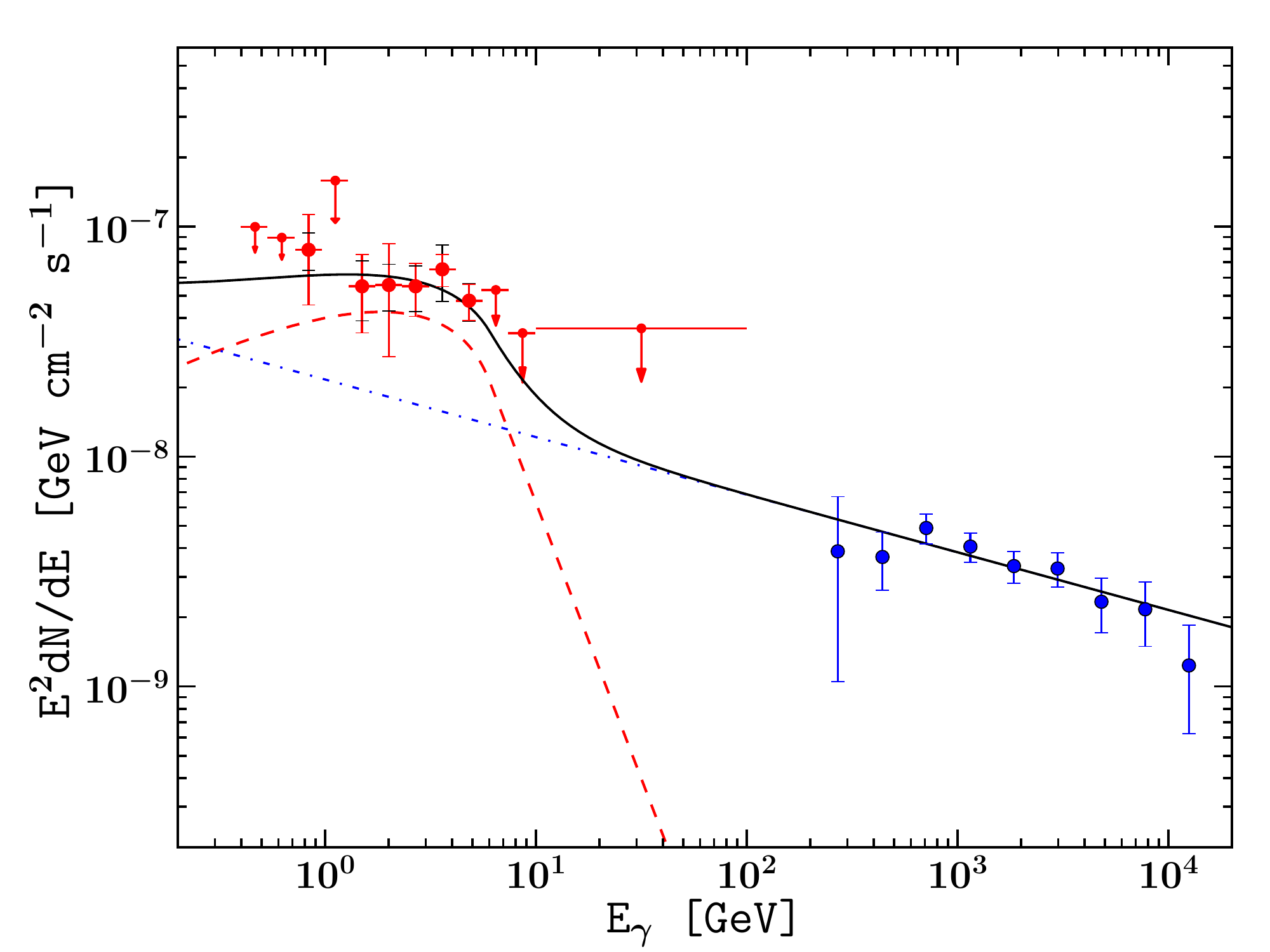} & 
\includegraphics[width=0.47\linewidth]{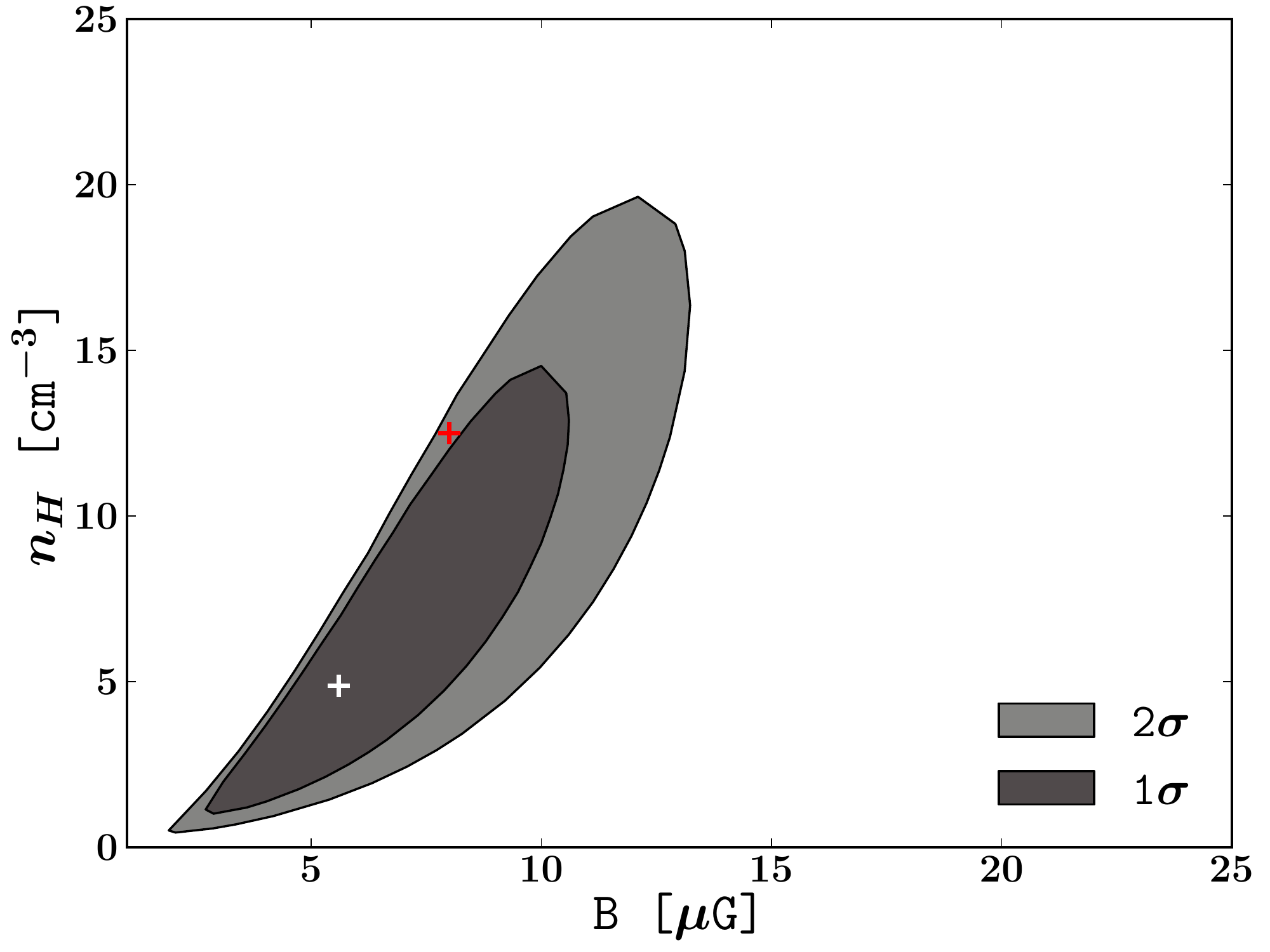} 
\end{tabular}

\caption{ \label{fig:HESSRidge}
\textit{(Left)} Red filled circles show the Fermi-LAT Galactic Ridge energy flux points obtained under the assumption of Model 2 in Table \ref{tab:LogLikelihoods}. They are listed in Table~\ref{tab:Ridgedata20-cmMapNOArcandSgrB}. The blue circles represent the Galactic Ridge as measured by the HESS telescope~\cite{Aharonian:2006au}. Black and red error bars show statistical and systematic errors respectively. Red arrows show $2\sigma$ upper limits. The red dashed curve is the gamma-ray nonthermal bremsstrahlung model generated from Eq.~\ref{eq:bremsstrahlungSpectra}. The blue dotted line
is a nonthermal bremsstrahlung model represented by a power law. The black solid line is the sum of the red dashed curve and blue dotted line.
It gives the best fit to the combined Fermi-LAT and HESS Galactic Ridge data.
\textit{(Right)} Confidence regions 
generated from the data and models shown in the left panel.
The parameter $n_H$ is the number density of hydrogen nuclei and $B$ the magnetic field. The white cross shows our best-fit values while the red cross corresponds to the values found in Ref.~\citep{yusef-zadehhewittwardle2013}. See also Table~\ref{tab:BremsstrahlungTable}.  }
\end{center}
\end{figure*}

Fitting the bremsstrahlung model (Eq.~\ref{eq:bremsstrahlungSpectra}), we varied the  number density of hydrogen nuclei $n_H$ and the magnetic field $B$. We simultaneously fit the normalization and slope of the power-law formula corresponding to the TeV HESS data. 
Using a bin-by-bin analysis we made a parameter scan
as shown in Fig.~\ref{fig:HESSRidge} and Table~\ref{tab:BremsstrahlungTable}. 

\begin{table*}[!t]
\begin{ruledtabular}
\begin{tabular}{lcccc}
\centering
 $n_{\rm H}$ [cm$^{-3}$] & $B$ [$\mu$G]& $N_0$ [ph cm$^{-2}$ s$^{-1}$ MeV$^{-1}$]& $\Gamma$ \\  \hline \\ 

 $5^{+6}_{-3}$ &$6^{+3}_{-2}$ &  $(2\pm 1)\times 10^{-11}$ & $2.25^{+0.07}_{-0.08}$\\

\end{tabular}
\end{ruledtabular}

\caption{\label{tab:BremsstrahlungTable} Best fit values obtained in the bremsstrahlung analysis for the gas number density ($n_{\rm H}$), the magnetic field ($B$), and the HESS power law spectrum amplitude ($N_0$) and spectral index ($\Gamma$). The best fit spectra and data fitted to are shown in the LHS panel of Fig.~ \ref{fig:HESSRidge}.}
\end{table*}

\begin{figure*}[ht!]
\begin{center}

%
%
\includegraphics[width=0.7\linewidth]{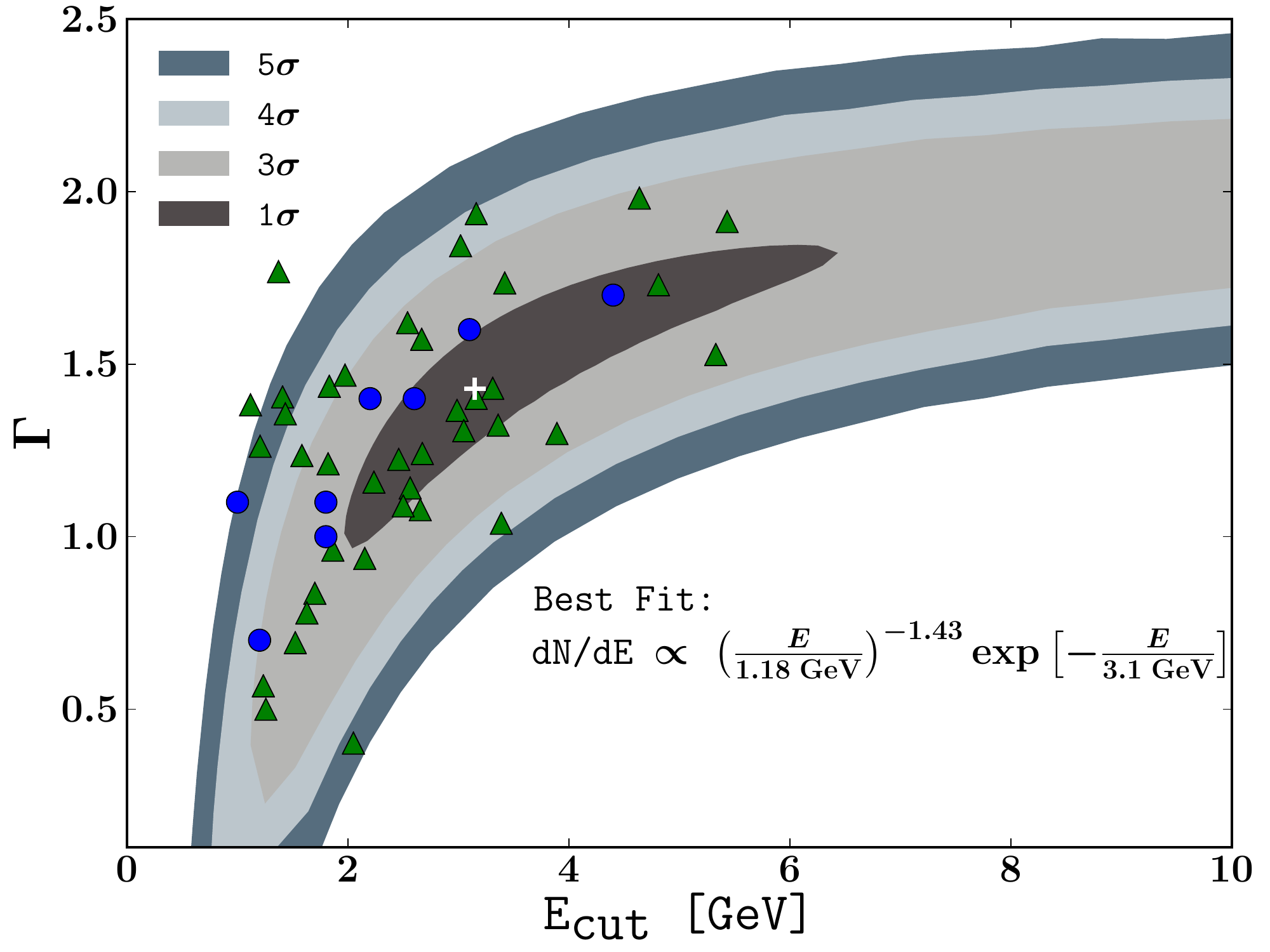} 

\caption{ \label{fig:PulsarsCL} Confidence  regions for an unresolved population of MSPs when the Galactic Ridge was included in the fit. 
 The data used is shown in the top RHS panel of Fig.~\ref{fig:galpropuncertainties} and listed in Table~\ref{tab:DMdata20-cmMap}.
 The best fit is denoted by a white cross.  The green triangles show the best fit parameters of the MSPs detected in the second Fermi LAT catalog of gamma-ray pulsars (2FPC)~\citep{FermiPulsarsCatalog}. The blue circles represent the best fit parameters of  MSP populated globular clusters~\citep{FermiGlobularClusters}.    }
\end{center}
\end{figure*}

\begin{figure*}[ht!]
\begin{center}

\begin{tabular}{cc}
\centering
\includegraphics[width=0.5\linewidth]{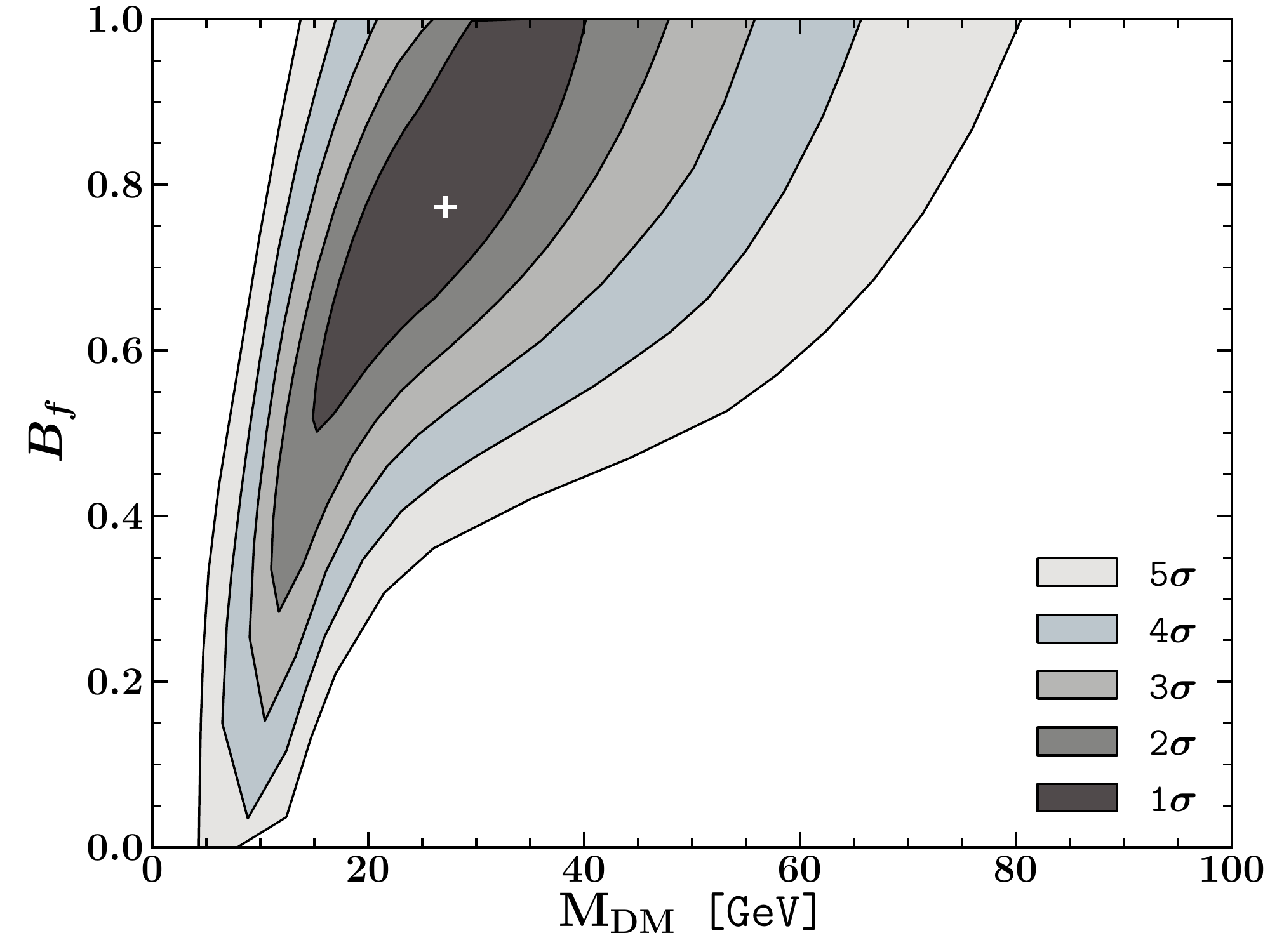} & \includegraphics[width=0.5\linewidth]{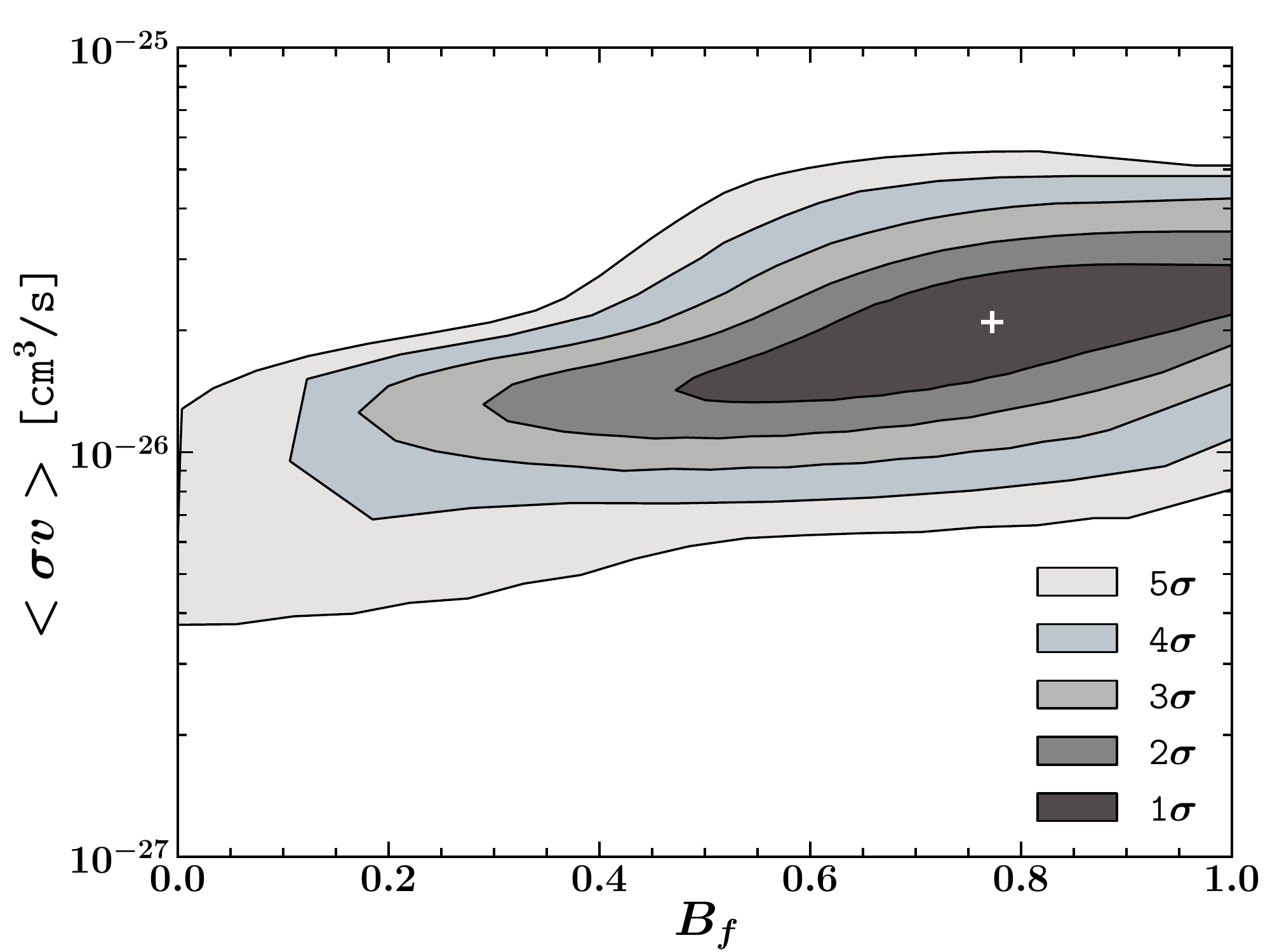}
\end{tabular}

\begin{tabular}{c}
\centering
 \includegraphics[width=0.7\linewidth]{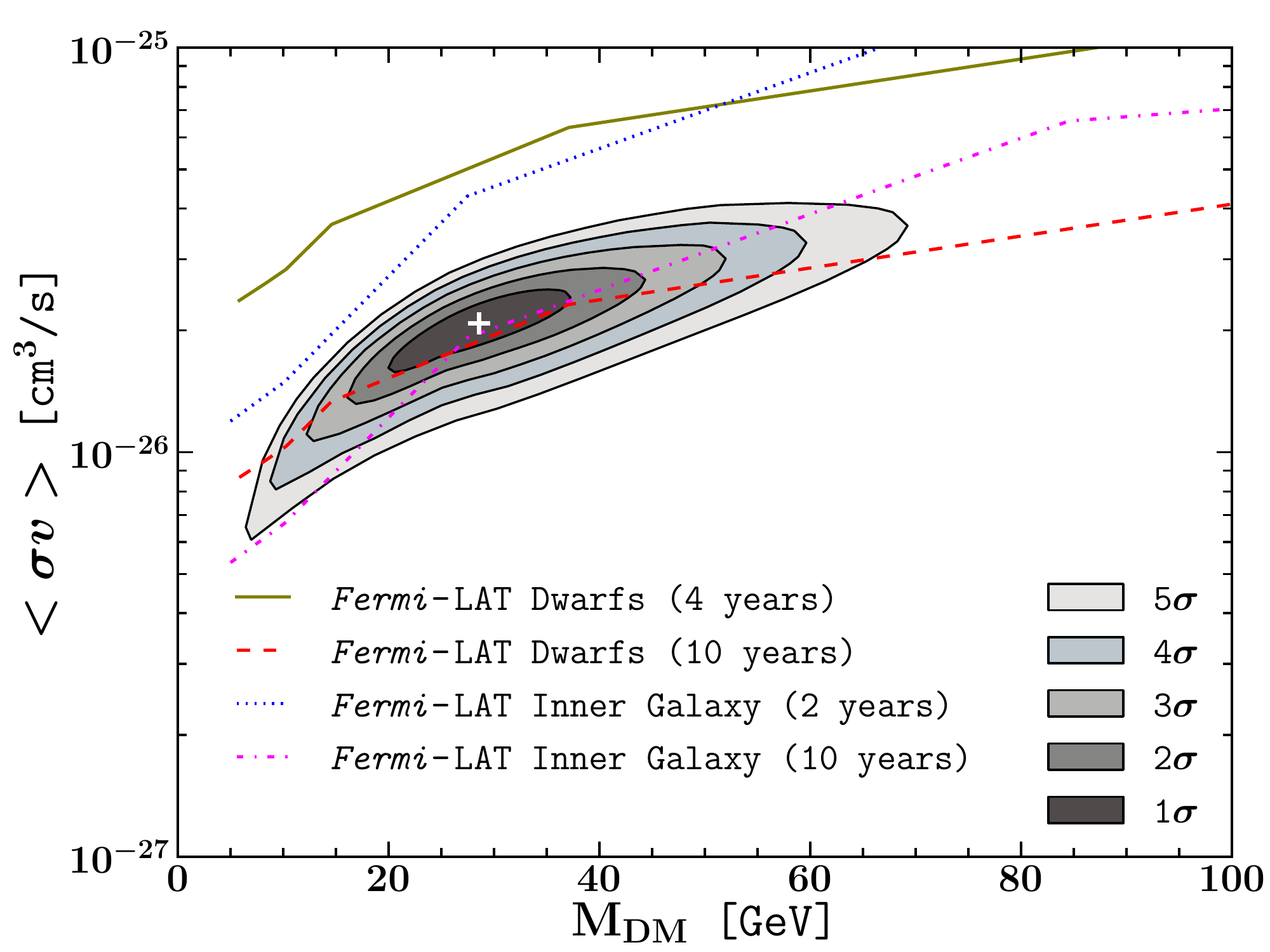}  

\end{tabular}

\caption{ \label{fig:darkmatterCL} Confidence regions for the dark matter model  when the Galactic Ridge source was included. 
The data used is shown in the top RHS panel of Fig.~\ref{fig:galpropuncertainties} and listed in Table~\ref{tab:DMdata20-cmMap}.
The parameter $B_f=1.0$ implies 100\% $b\bar{b}$ and $B_f=0.0$ means 100\% $\tau^+\tau^-$. The DM spatial distribution follows a NFW profile with inner slope $\gamma=1.2$. The white cross denotes the best-fit values. Limits from dwarf galaxies \citep{FermiDwarf4yr} (2$\sigma$) and the Inner Galaxy ($10^\circ\leq b\leq 20^\circ$) (3$\sigma$) \citep{FermiInnerGalaxy2012} are included. We rescaled the Inner Galaxy results to account for the different $\gamma$ of the current study, using the Galactic coordinate ($l=0,b=10$) as a reference point \citep{GordonMacias2013}. The 10 year forecasts were approximated with a simple $1/\sqrt{\rm time}$ scaling  and in the dwarf galaxy case it was assumed there were three times more dwarf galaxies available.}
\end{center}
\end{figure*}

Additionally, this analysis enabled us to study to what extent the Galactic Ridge component affects the model parameters of a DM or unresolved MSPs extended source. We therefore made a detailed parameter scan corresponding to the DM and MSPs hypotheses in models which included a Galactic Ridge. The dark matter spectra are obtained using DMFIT~\citep{Profumo1} while the standard exponential cut off form is used for the MSPs' spectrum:
\begin{equation}
\frac{dN}{dE}=K\left({E\over E_0}\right)^{-\Gamma}\exp\left(-\frac{E}{E_{\rm cut}}\right),
\label{eq:expcut}
\end{equation}    
where photon index $\Gamma$, a cut-off energy $E_{\rm cut}$ and a normalization factor $K$ are free parameters. 
 The results are summarized in Fig.~\ref{fig:PulsarsCL} and Table~\ref{tab:Pulsars} for the MSPs hypothesis, and Fig.~\ref{fig:darkmatterCL} and Table~\ref{tab:darkmatter} for the DM hypothesis.

\begin{table*}[!t]

\begin{ruledtabular}
\begin{tabular}{lccc}
\centering
 Model&$E_{\rm cut}$ [GeV] & $\Gamma$  &$G_{100}$ [$10^{-9}$ erg cm$^{-2}$ s$^{-1}$]\\  \hline \\ 
MSPs    &$4^{+2}_{-1}$ & $1.6\pm 0.2$ & $1.5\pm 0.2$ \\
 MSPs $+$ Galactic Ridge   & $3^{+2}_{-1}$ & $1.4\pm 0.3$ & $1.2^{+0.2}_{-0.1}$

\end{tabular}
\end{ruledtabular}

\caption{\label{tab:Pulsars} Best-fit values for MSPs hypothesis. 
The spectrum of the MSPs is fitted with a power law with an exponential cut off (see Fig.~\ref{fig:PulsarsCL}).  The first row shows the result from an analysis without a galactic ridge \cite{GordonMacias2013}. The second row parameters were fitted to the spectral data plotted on the top RHS panel of Fig.~\ref{fig:galpropuncertainties}. The GCEG energy flux for $100$ MeV$\leq E\leq 100$ GeV is denoted by $G_{100}$.}
\end{table*}

\begin{table*}[!t]

\begin{ruledtabular}
\begin{tabular}{lccc}
\centering
Model&Best-fit Branching ratio & $\left<\sigma v\right>$ [cm$^3$/s] & $M_{\rm DM}$ [GeV] \\  \hline \\ 
DM &$(60\pm20)\%$ $b\bar{b}$    & $(2.8\pm0.4)\times 10^{-26}$ & $24\pm 7$\\
DM $+$ 20-cm template& $(80\pm20)\%$ $b\bar{b}$    & $2.0^{+0.5}_{-0.6}\times 10^{-26}$ & $27^{+8}_{-9}$ 
\end{tabular}
\end{ruledtabular}

\caption{\label{tab:darkmatter} Best-fit values for the branching fraction between $b\bar{b}$ and $\tau^+\tau^-$, DM velocity averaged annihilation cross section and DM mass. DMFIT was used to generate the model spectra~\citep{Profumo1}.
The first row shows the result from an analysis without a Galactic 
Ridge  \cite{GordonMacias2013}. The second row parameters were fitted to the spectral data plotted on the top RHS panel of Fig.~\ref{fig:galpropuncertainties}.
}
\end{table*}

\section{Discussion}
\label{sec:Discussion}
The main focus of our study was try to evaluate three competing explanations for the GCEG: MSPs, DM, or a Galactic Ridge resulting from the interaction of cosmic rays with molecular clouds. As we discuss below, the data prefer combinations of the Galactic Ridge template and a NFW$_{1.2}^2$ template which has a spectrum compatible with either MSPs, DM, or some combination of the two. 

\subsection{Interaction of cosmic rays with molecular clouds}

From Table~\ref{tab:LogLikelihoods} we 
can check the significance of adding a new component by evaluating 
the test statistic (TS) which is defined as in Ref.~\cite{2FGL}:
\begin{eqnarray}\label{tsdef}
{\rm TS}=2\left[\log \mathcal{L} (\mbox{new source})-\log \mathcal{L} (\mbox{NO-new source})\right]\mbox{,} \qquad 
\end{eqnarray}
where $\mathcal{L}$ stands for the maximum of the likelihood of the data given the model with or without the new source.
In the large sample limit, under the no source hypothesis, TS has a $\chi^2/2$ distribution with the number of degrees of freedom equal to the number of 
parameters associated with the proposed positive amplitude new source  \cite{wilks,mattox}. As the amplitude is restricted to be non-negative, a $\chi^2/2$ distribution rather than the $\chi^2$ distribution is needed.

As can be seen from Table~\ref{tab:LogLikelihoods}, the improvement in the fit of the 20-cm Galactic Ridge relative to 2FGL is TS$=648-425=213$ for $13-9=4$ extra degrees of freedom (dof). This corresponds to a $14\sigma$ detection (if we convert to the equivalent p-value for 1 degree of freedom) and so confirms that the 20-cm Galactic Ridge does improve the fit to the GCEG. However, the corresponding TS for a NFW$_{1.2}^2$ template is 870 and for only 3 extra dof and so clearly also improves the fit substantially. 

We can check whether the 20-cm Galactic Ridge still improves the fit once the        NFW$_{1.2}^2$ template is included. 
From Table~\ref{tab:LogLikelihoods} we obtain a TS$=1330-1295=35$ for 4 extra dof which corresponds to a 5$\sigma$ detection. This shows that the GCEG motivates a sum of the NFW$_{1.2}^2$ and the Galactic Ridge being included.

The parts of the data which require the NFW$_{1.2}^2$ and the 20-cm Galactic Ridge are shown in Fig.~\ref{fig:residuals}.  The elongation in the longitudinal direction, indicating the need for the 20-cm Galactic Ridge, is particularly evident in the energy ranges 1.73 to 5.57 GeV.

We also did the above analysis with the HESS residual Galactic Ridge and we found that a TS=30 for 4 extra dof  
which is less than the 20-cm case, but the difference is not statistically significant.

Additionally, we checked whether the inclusion of the  Galactic Ridge affected the spectral parameters of the NFW$_{1.2}^2$ model. As can be seen from Tables~\ref{tab:Pulsars} and \ref{tab:darkmatter}, the inclusion of the Galactic Ridge does not significantly alter the spectral parameters of the NFW$_{1.2}^2$ template.

In Ref.~\cite{yusef-zadehhewittwardle2013}, it was argued that the Arc and Sgr B were associated with cosmic rays interacting with molecular clouds and so should not be included when evaluating the parameters of the Galactic Ridge. They also investigated the effects of adding Sgr C, but we found once the NFW$_{1.2}^2$ was included, Sgr C had a very low TS and so we have not included it in our analysis.

As can be seen from Table~\ref{tab:LogLikelihoods}, the Arc and Sgr B do significantly improve the fit even when the Galactic Ridge and the NFW$_{1.2}^2$ template are included. However, it is common practice \cite{Uchiyama&Funk,Ackermann:2013wqa} to exclude such point sources  when analyzing a physical model for the cosmic rays interacting with molecular gas. Otherwise, some of the signal will be lost to the apparent  point sources  arising from cosmic rays interacting with molecular gas. As shown in Table~\ref{tab:LogLikelihoods}, in this case there is also no significant difference in the goodness of fit between HESS Galactic Ridge and 20-cm Galactic Ridge, but as the 20-cm template has  a slightly higher TS in both models 1 and 2, we use it as the default.

In Fig.~\ref{fig:HESSRidge} we provide confidence regions for the magnetic field $B$ and hydrogen density $n_H$.
 The Galactic Ridge is consistent with the Ref.~\cite{yusef-zadehhewittwardle2013}  best fit even though the extra NFW$_{1.2}^2$ component is included.

\subsection{Millisecond Pulsars}
In Fig.~\ref{fig:PulsarsCL} we show the confidence intervals for the exponential cut off fit. Although, they are not significantly different from the ones without an extended Galactic Ridge template (see Ref.\cite{GordonMacias2013}), here we also show the MSPs reported in the second Fermi LAT catalog of gamma-ray pulsars (2FPC)~\cite{FermiPulsarsCatalog} and also the globular clusters which can contain multiple unresolved MSPs~\cite{FermiGlobularClusters}. As can be seen the spectrum of the GCEG is consistent with the majority of MSPs and MSP containing globular clusters.

Using the GCEG energy flux for $100$ MeV$\leq E\leq 100$ GeV ($G_{100}$) of our best fit exponential cut off model from Table \ref{tab:Pulsars}
we evaluate the luminosity as $L_\gamma=4\pi d^2 G_{100}\sim 10^{37}$ erg s$^{-1}$ where we take the distance to the Galactic Center as $d\sim 8$ kpc. 

The 2FPC 
contains 40 MSPs with estimated luminosities ranging from about 
$5\times 10^{31}$  to $7\times 10^{34}$ erg s$^{-1}$. The average MSP luminosity in the 2FPC  is  $\bar{L}_{\rm MSP} \sim10^{34}$ erg s$^{-1}$. Only about 20\% of known MSPs have been detected by Fermi-LAT \cite{FermiPulsarsCatalog} and so the catalog is biased towards higher gamma-ray luminosity MSPs. Therefore, we expect the 2FPC average MSP luminosity will be greater than the MSP population average. So we use the average 2FPC value to estimate a lower bound of  $\sim$ 1000 MSPs  for $r\lesssim 150$pc in order to explain the GCEG.

If we assume each of the MSPs at the Galactic Center has a luminosity of $\bar{L}_{\rm MSP}$ and then convert this to a flux using $d\sim 8$ kpc we get each MSP at the Galactic Center has a flux $\sim 10^{-12}$ erg cm $^{-2}$ s$^{-1}$ which, as can be seen from Fig.~17 of Ref.~\cite{FermiPulsarsCatalog}, is below the detection limit ($\sim  10^{-11}$ erg cm $^{-2}$ s$^{-1}$) at the Galactic Center. This is consistent with these proposed Galactic Center MSPs being unresolved.

The Galactic Center $r\lesssim 150$ pc region corresponds to about 6 square degrees. As the Fermi-LAT resolution is $\gtrsim0.1^\circ$ at the relevant energy level of this work, it follows that each ($0.1$ deg)$^2$ pixel of the Fermi-LAT image of the Galactic Center would contain $\gtrsim 1$ MSP.

This MSPs explanation of the GCEG is consistent with the results presented in Ref.~\cite{wharton2012}. Their analysis is based on the number of neutron stars which are computed from the core collapse supernovae rate which in turn is obtained from measurements of the total mass of $^{26}$Al in the Galaxy. Using this method they estimate the number of MSPs as $\sim 10^5 f_r$ for $r\lesssim 150$ pc where $f_r$ is the fraction of neutron stars that get recycled to MSPs. Based on Galactic disk and globular cluster radio observations, they estimate $f_r\sim10^{-2}$ for $r\lesssim 150$ pc.

A justification for the MSPs resulting in a NFW$_{1.2}^2$ profile with  $\gamma\sim 1.2$ was proposed in Ref.~\cite{AK} who noted the observations of low mass X-ray binaries (LMXBs) in M31 follow a similar profile and MSPs are believed to arise from LMXBs. They also note some indication of a similar trend in the Milky Way although the LMXB observational data in that case is currently not very conclusive.

\subsection{Dark Matter}
Although the estimates for the DM parameters are not significantly changed, as can be seen from Fig.~\ref{fig:darkmatterCL}, the $\tau^+\tau^-$ channel is now only excluded as 4$\sigma$ rather than 5$\sigma$ as was the case when no Galactic Ridge was included \cite{GordonMacias2013}.  As can be seen in the bottom panel of Fig.~\ref{fig:darkmatterCL}, neither Fermi-LAT dwarfs nor the Fermi-LAT inner Galaxy will be able to definitively detect the DM self-annihilation if it is causing the GCEG. Also, as there is likely to be at least some MSP contribution, the actual $\left< \sigma v \right>$ will be correspondingly lower and so even harder to detect.  

\section{Conclusions}
We have found that the GCEG is best fit by adding to the base 2FGL model both a NFW$_{1.2}^2$ source and a Galactic Ridge based on a 20-cm continuum emission template. Similar results were found for a Galactic Ridge template based on the HESS data residuals. The addition of the Galactic Ridge was not found to significantly affect the NFW$_{1.2}^2$ spectral parameters. We found that the GCEG is consistent with a lower bound of $\sim 1000$ on the number of MSPs at the Galactic Center. This is consistent with estimates based on core collapse supernovae inferences from $^{26}$Al measurements. We also demonstrated that current and 10-year Fermi-LAT measurements of dwarf spheroidals and the inner Galaxy are unlikely to be able to conclusively check a  DM annihilation explanation of the GCEG. 
As the modeling based on the $^{26}$Al measurements indicates there is likely to be $\sim 1000 $ MSPs in the Galactic Center, this implies that if there is a DM annihilation component then it probably has a significantly smaller $\left< \sigma v \right>$ and so will be even harder to check.

We also constrained a bremsstrahlung model of the Galactic Ridge and showed that
 the $B$ and $n_H$ constraints are consistent with a previous analysis~\citep{yusef-zadehhewittwardle2013} done without a NFW$_{1.2}^2$ component. 

\begin{acknowledgements}
We thank Mark Wardle for providing notes for the bremsstrahlung formula used in the analysis of Ref.~\citep{yusef-zadehhewittwardle2013}. We also thank Farhad Yusef-Zadeh for providing the 20-cm template and helpful comments on the underlying physics it measures.
O.M. is supported by a UC Doctoral Scholarship. This work makes use of \textsc{Fermi Science Tools}~\citep{fermitools}, \textsc{DMFit}~\citep{Profumo1}, \textsc{Minuit}~\citep{minuit},  \textsc{SciPy}~\citep{scipy}, and GALPROP~\citep{GALPROP, webrun}.

\end{acknowledgements}

\bibliography{references}

\begin{thebibliography}{46}%
\makeatletter
\providecommand \@ifxundefined [1]{%
 \@ifx{#1\undefined}
}%
\providecommand \@ifnum [1]{%
 \ifnum #1\expandafter \@firstoftwo
 \else \expandafter \@secondoftwo
 \fi
}%
\providecommand \@ifx [1]{%
 \ifx #1\expandafter \@firstoftwo
 \else \expandafter \@secondoftwo
 \fi
}%
\providecommand \natexlab [1]{#1}%
\providecommand \enquote  [1]{``#1''}%
\providecommand \bibnamefont  [1]{#1}%
\providecommand \bibfnamefont [1]{#1}%
\providecommand \citenamefont [1]{#1}%
\providecommand \href@noop [0]{\@secondoftwo}%
\providecommand \href [0]{\begingroup \@sanitize@url \@href}%
\providecommand \@href[1]{\@@startlink{#1}\@@href}%
\providecommand \@@href[1]{\endgroup#1\@@endlink}%
\providecommand \@sanitize@url [0]{\catcode `\\12\catcode `\$12\catcode
  `\&12\catcode `\#12\catcode `\^12\catcode `\_12\catcode `\%12\relax}%
\providecommand \@@startlink[1]{}%
\providecommand \@@endlink[0]{}%
\providecommand \url  [0]{\begingroup\@sanitize@url \@url }%
\providecommand \@url [1]{\endgroup\@href {#1}{\urlprefix }}%
\providecommand \urlprefix  [0]{URL }%
\providecommand \Eprint [0]{\href }%
\providecommand \doibase [0]{http://dx.doi.org/}%
\providecommand \selectlanguage [0]{\@gobble}%
\providecommand \bibinfo  [0]{\@secondoftwo}%
\providecommand \bibfield  [0]{\@secondoftwo}%
\providecommand \translation [1]{[#1]}%
\providecommand \BibitemOpen [0]{}%
\providecommand \bibitemStop [0]{}%
\providecommand \bibitemNoStop [0]{.\EOS\space}%
\providecommand \EOS [0]{\spacefactor3000\relax}%
\providecommand \BibitemShut  [1]{\csname bibitem#1\endcsname}%
\let\auto@bib@innerbib\@empty
\bibitem [{\citenamefont {Cirelli}(2012)}]{cirelli2}%
  \BibitemOpen
  \bibfield  {author} {\bibinfo {author} {\bibfnamefont {Marco}\ \bibnamefont
  {Cirelli}},\ }\bibfield  {title} {\enquote {\bibinfo {title} {{Indirect
  Searches for Dark Matter: a status review}},}\ }\href@noop {} {\  (\bibinfo
  {year} {2012})},\ \Eprint {http://arxiv.org/abs/1202.1454} {arXiv:1202.1454
  [hep-ph]} \BibitemShut {NoStop}%
\bibitem [{\citenamefont {Funk}(2013)}]{Funk}%
  \BibitemOpen
  \bibfield  {author} {\bibinfo {author} {\bibfnamefont {Stefan}\ \bibnamefont
  {Funk}},\ }\bibfield  {title} {\enquote {\bibinfo {title} {{Indirect
  Detection of Dark Matter with gamma rays}},}\ }\href@noop {} {\  (\bibinfo
  {year} {2013})},\ \Eprint {http://arxiv.org/abs/1310.2695} {arXiv:1310.2695
  [astro-ph.HE]} \BibitemShut {NoStop}%
\bibitem [{\citenamefont {Goodenough}\ and\ \citenamefont
  {Hooper}(2009)}]{Goodenough:2009gk}%
  \BibitemOpen
  \bibfield  {author} {\bibinfo {author} {\bibfnamefont {Lisa}\ \bibnamefont
  {Goodenough}}\ and\ \bibinfo {author} {\bibfnamefont {Dan}\ \bibnamefont
  {Hooper}},\ }\bibfield  {title} {\enquote {\bibinfo {title} {{Possible
  Evidence For Dark Matter Annihilation In The Inner Milky Way From The Fermi
  Gamma Ray Space Telescope}},}\ }\href@noop {} {\  (\bibinfo {year} {2009})},\
  \Eprint {http://arxiv.org/abs/0910.2998} {arXiv:0910.2998 [hep-ph]}
  \BibitemShut {NoStop}%
\bibitem [{\citenamefont {Hooper}\ and\ \citenamefont
  {Goodenough}(2011)}]{Hooper:2010mq}%
  \BibitemOpen
  \bibfield  {author} {\bibinfo {author} {\bibfnamefont {D.}~\bibnamefont
  {Hooper}}\ and\ \bibinfo {author} {\bibfnamefont {L.}~\bibnamefont
  {Goodenough}},\ }\bibfield  {title} {\enquote {\bibinfo {title} {{Dark Matter
  Annihilation in The Galactic Center As Seen by the Fermi Gamma Ray Space
  Telescope}},}\ }\href {\doibase 10.1016/j.physletb.2011.02.029} {\bibfield
  {journal} {\bibinfo  {journal} {Phys.Lett.}\ }\textbf {\bibinfo {volume}
  {B697}},\ \bibinfo {pages} {412--428} (\bibinfo {year} {2011})},\ \Eprint
  {http://arxiv.org/abs/1010.2752} {arXiv:1010.2752 [hep-ph]} \BibitemShut
  {NoStop}%
\bibitem [{\citenamefont {Boyarsky}\ \emph {et~al.}(2011)\citenamefont
  {Boyarsky}, \citenamefont {Malyshev},\ and\ \citenamefont
  {Ruchayskiy}}]{Boyarsky:2010dr}%
  \BibitemOpen
  \bibfield  {author} {\bibinfo {author} {\bibfnamefont {Alexey}\ \bibnamefont
  {Boyarsky}}, \bibinfo {author} {\bibfnamefont {Denys}\ \bibnamefont
  {Malyshev}}, \ and\ \bibinfo {author} {\bibfnamefont {Oleg}\ \bibnamefont
  {Ruchayskiy}},\ }\bibfield  {title} {\enquote {\bibinfo {title} {{A comment
  on the emission from the Galactic Center as seen by the Fermi telescope}},}\
  }\href {\doibase 10.1016/j.physletb.2011.10.014} {\bibfield  {journal}
  {\bibinfo  {journal} {Phys.Lett.}\ }\textbf {\bibinfo {volume} {B705}},\
  \bibinfo {pages} {165--169} (\bibinfo {year} {2011})},\ \Eprint
  {http://arxiv.org/abs/1012.5839} {arXiv:1012.5839 [hep-ph]} \BibitemShut
  {NoStop}%
\bibitem [{\citenamefont {Hooper}\ and\ \citenamefont
  {Linden}(2011)}]{hooperlinden2011}%
  \BibitemOpen
  \bibfield  {author} {\bibinfo {author} {\bibfnamefont {D.}~\bibnamefont
  {Hooper}}\ and\ \bibinfo {author} {\bibfnamefont {T.}~\bibnamefont
  {Linden}},\ }\bibfield  {title} {\enquote {\bibinfo {title} {{On The Origin
  Of The Gamma Rays From The Galactic Center}},}\ }\href {\doibase
  10.1103/PhysRevD.84.123005} {\bibfield  {journal} {\bibinfo  {journal}
  {Phys.Rev.}\ }\textbf {\bibinfo {volume} {D84}},\ \bibinfo {pages} {123005}
  (\bibinfo {year} {2011})},\ \Eprint {http://arxiv.org/abs/1110.0006}
  {arXiv:1110.0006 [astro-ph.HE]} \BibitemShut {NoStop}%
\bibitem [{\citenamefont {Hooper}\ \emph {et~al.}(2012)\citenamefont {Hooper},
  \citenamefont {Kelso},\ and\ \citenamefont
  {Queiroz}}]{hooperkelsoqueiroz2012}%
  \BibitemOpen
  \bibfield  {author} {\bibinfo {author} {\bibfnamefont {Dan}\ \bibnamefont
  {Hooper}}, \bibinfo {author} {\bibfnamefont {Chris}\ \bibnamefont {Kelso}}, \
  and\ \bibinfo {author} {\bibfnamefont {Farinaldo~S.}\ \bibnamefont
  {Queiroz}},\ }\bibfield  {title} {\enquote {\bibinfo {title} {{Stringent and
  Robust Constraints on the Dark Matter Annihilation Cross Section From the
  Region of the Galactic Center}},}\ }\href@noop {} {\  (\bibinfo {year}
  {2012})},\ \Eprint {http://arxiv.org/abs/1209.3015} {arXiv:1209.3015
  [astro-ph.HE]} \BibitemShut {NoStop}%
\bibitem [{\citenamefont {Abazajian}\ and\ \citenamefont
  {Kaplinghat}(2012)}]{AK}%
  \BibitemOpen
  \bibfield  {author} {\bibinfo {author} {\bibfnamefont {Kevork~N.}\
  \bibnamefont {Abazajian}}\ and\ \bibinfo {author} {\bibfnamefont {Manoj}\
  \bibnamefont {Kaplinghat}},\ }\bibfield  {title} {\enquote {\bibinfo {title}
  {{Detection of a Gamma-Ray Source in the Galactic Center Consistent with
  Extended Emission from Dark Matter Annihilation and Concentrated
  Astrophysical Emission}},}\ }\href {\doibase 10.1103/PhysRevD.86.083511}
  {\bibfield  {journal} {\bibinfo  {journal} {Phys.Rev.}\ }\textbf {\bibinfo
  {volume} {D86}},\ \bibinfo {pages} {083511} (\bibinfo {year} {2012})},\
  \Eprint {http://arxiv.org/abs/1207.6047} {arXiv:1207.6047 [astro-ph.HE]}
  \BibitemShut {NoStop}%
\bibitem [{\citenamefont {Abazajian}\ and\ \citenamefont
  {Kaplinghat}(2013)}]{AKerratum}%
  \BibitemOpen
  \bibfield  {author} {\bibinfo {author} {\bibfnamefont {Kevork~N.}\
  \bibnamefont {Abazajian}}\ and\ \bibinfo {author} {\bibfnamefont {Manoj}\
  \bibnamefont {Kaplinghat}},\ }\bibfield  {title} {\enquote {\bibinfo {title}
  {Erratum: Detection of a gamma-ray source in the galactic center consistent
  with extended emission from dark matter annihilation and concentrated
  astrophysical emission [phys. rev. d 86, 083511 (2012)]},}\ }\href {\doibase
  10.1103/PhysRevD.87.129902} {\bibfield  {journal} {\bibinfo  {journal} {Phys.
  Rev. D}\ }\textbf {\bibinfo {volume} {87}},\ \bibinfo {pages} {129902}
  (\bibinfo {year} {2013})}\BibitemShut {NoStop}%
\bibitem [{\citenamefont {Gordon}\ and\ \citenamefont
  {Macias}(2013)}]{GordonMacias2013}%
  \BibitemOpen
  \bibfield  {author} {\bibinfo {author} {\bibfnamefont {Chris}\ \bibnamefont
  {Gordon}}\ and\ \bibinfo {author} {\bibfnamefont {Oscar}\ \bibnamefont
  {Macias}},\ }\bibfield  {title} {\enquote {\bibinfo {title} {{Dark Matter and
  Pulsar Model Constraints from Galactic Center Fermi-LAT Gamma Ray
  Observations}},}\ }\href@noop {} {\  (\bibinfo {year} {2013})},\ \Eprint
  {http://arxiv.org/abs/1306.5725} {arXiv:1306.5725 [astro-ph.HE]} \BibitemShut
  {NoStop}%
\bibitem [{\citenamefont {Vitale}\ and\ \citenamefont
  {Morselli}(2009)}]{Vitale:2009hr}%
  \BibitemOpen
  \bibfield  {author} {\bibinfo {author} {\bibfnamefont {Vincenzo}\
  \bibnamefont {Vitale}}\ and\ \bibinfo {author} {\bibfnamefont {Aldo}\
  \bibnamefont {Morselli}} (\bibinfo {collaboration} {Fermi/LAT
  Collaboration}),\ }\bibfield  {title} {\enquote {\bibinfo {title} {{Indirect
  Search for Dark Matter from the center of the Milky Way with the Fermi-Large
  Area Telescope}},}\ }\href@noop {} {\  (\bibinfo {year} {2009})},\ \Eprint
  {http://arxiv.org/abs/0912.3828} {arXiv:0912.3828 [astro-ph.HE]} \BibitemShut
  {NoStop}%
\bibitem [{\citenamefont {{Vitale}}\ \emph {et~al.}(2011)\citenamefont
  {{Vitale}}, \citenamefont {{Morselli}},\ and\ \citenamefont {{Fermi/LAT
  Collaboration}}}]{2011NIMPA.630..147V}%
  \BibitemOpen
  \bibfield  {author} {\bibinfo {author} {\bibfnamefont {V.}~\bibnamefont
  {{Vitale}}}, \bibinfo {author} {\bibfnamefont {A.}~\bibnamefont
  {{Morselli}}}, \ and\ \bibinfo {author} {\bibnamefont {{Fermi/LAT
  Collaboration}}},\ }\bibfield  {title} {\enquote {\bibinfo {title} {{Search
  for Dark Matter with Fermi Large Area Telescope: The Galactic Center}},}\
  }\href {\doibase 10.1016/j.nima.2010.06.048} {\bibfield  {journal} {\bibinfo
  {journal} {Nuclear Instruments and Methods in Physics Research A}\ }\textbf
  {\bibinfo {volume} {630}},\ \bibinfo {pages} {147--150} (\bibinfo {year}
  {2011})}\BibitemShut {NoStop}%
\bibitem [{\citenamefont {Iocco}\ \emph {et~al.}(2011)\citenamefont {Iocco},
  \citenamefont {Pato}, \citenamefont {Bertone},\ and\ \citenamefont
  {Jetzer}}]{ioccopatobertone2011}%
  \BibitemOpen
  \bibfield  {author} {\bibinfo {author} {\bibfnamefont {Fabio}\ \bibnamefont
  {Iocco}}, \bibinfo {author} {\bibfnamefont {Miguel}\ \bibnamefont {Pato}},
  \bibinfo {author} {\bibfnamefont {Gianfranco}\ \bibnamefont {Bertone}}, \
  and\ \bibinfo {author} {\bibfnamefont {Philippe}\ \bibnamefont {Jetzer}},\
  }\bibfield  {title} {\enquote {\bibinfo {title} {{Dark Matter distribution in
  the Milky Way: microlensing and dynamical constraints}},}\ }\href {\doibase
  10.1088/1475-7516/2011/11/029} {\bibfield  {journal} {\bibinfo  {journal}
  {JCAP}\ }\textbf {\bibinfo {volume} {1111}},\ \bibinfo {pages} {029}
  (\bibinfo {year} {2011})},\ \Eprint {http://arxiv.org/abs/1107.5810}
  {arXiv:1107.5810 [astro-ph.GA]} \BibitemShut {NoStop}%
\bibitem [{\citenamefont {Abazajian}(2011)}]{Abazajian:2010zy}%
  \BibitemOpen
  \bibfield  {author} {\bibinfo {author} {\bibfnamefont {Kevork~N.}\
  \bibnamefont {Abazajian}},\ }\bibfield  {title} {\enquote {\bibinfo {title}
  {{The Consistency of Fermi-LAT Observations of the Galactic Center with a
  Millisecond Pulsar Population in the Central Stellar Cluster}},}\ }\href
  {\doibase 10.1088/1475-7516/2011/03/010} {\bibfield  {journal} {\bibinfo
  {journal} {JCAP}\ }\textbf {\bibinfo {volume} {1103}},\ \bibinfo {pages}
  {010} (\bibinfo {year} {2011})},\ \Eprint {http://arxiv.org/abs/1011.4275}
  {arXiv:1011.4275 [astro-ph.HE]} \BibitemShut {NoStop}%
\bibitem [{\citenamefont {{Wharton}}\ \emph {et~al.}(2012)\citenamefont
  {{Wharton}}, \citenamefont {{Chatterjee}}, \citenamefont {{Cordes}},
  \citenamefont {{Deneva}},\ and\ \citenamefont {{Lazio}}}]{wharton2012}%
  \BibitemOpen
  \bibfield  {author} {\bibinfo {author} {\bibfnamefont {R.~S.}\ \bibnamefont
  {{Wharton}}}, \bibinfo {author} {\bibfnamefont {S.}~\bibnamefont
  {{Chatterjee}}}, \bibinfo {author} {\bibfnamefont {J.~M.}\ \bibnamefont
  {{Cordes}}}, \bibinfo {author} {\bibfnamefont {J.~S.}\ \bibnamefont
  {{Deneva}}}, \ and\ \bibinfo {author} {\bibfnamefont {T.~J.~W.}\ \bibnamefont
  {{Lazio}}},\ }\bibfield  {title} {\enquote {\bibinfo {title}
  {{Multiwavelength Constraints on Pulsar Populations in the Galactic
  Center}},}\ }\href {\doibase 10.1088/0004-637X/753/2/108} {\bibfield
  {journal} {\bibinfo  {journal} {\apj}\ }\textbf {\bibinfo {volume} {753}},\
  \bibinfo {eid} {108} (\bibinfo {year} {2012})},\ \Eprint
  {http://arxiv.org/abs/1111.4216} {arXiv:1111.4216 [astro-ph.HE]} \BibitemShut
  {NoStop}%
\bibitem [{\citenamefont {Gordon}\ and\ \citenamefont
  {Macias}(2014)}]{GordonMacias2013erratum}%
  \BibitemOpen
  \bibfield  {author} {\bibinfo {author} {\bibfnamefont {Chris}\ \bibnamefont
  {Gordon}}\ and\ \bibinfo {author} {\bibfnamefont {Oscar}\ \bibnamefont
  {Macias}},\ }\bibfield  {title} {\enquote {\bibinfo {title} {{Erratum: Dark
  Matter and Pulsar Model Constraints from Galactic Center Fermi-LAT Gamma Ray
  Observations}},}\ }\href@noop {} {\bibfield  {journal} {\bibinfo  {journal}
  {Phys. Rev. D}\ } (\bibinfo {year} {2014})},\ \Eprint
  {http://arxiv.org/abs/1306.5725} {arXiv:1306.5725 [astro-ph.HE]} \BibitemShut
  {NoStop}%
\bibitem [{\citenamefont {{Mirabal}}(2013)}]{Miribal2013}%
  \BibitemOpen
  \bibfield  {author} {\bibinfo {author} {\bibfnamefont {N.}~\bibnamefont
  {{Mirabal}}},\ }\bibfield  {title} {\enquote {\bibinfo {title} {{Dark matter
  versus pulsars: catching the impostor}},}\ }\href {\doibase
  10.1093/mnras/stt1740} {\bibfield  {journal} {\bibinfo  {journal} {\mnras}\
  }\textbf {\bibinfo {volume} {436}},\ \bibinfo {pages} {2461--2464} (\bibinfo
  {year} {2013})},\ \Eprint {http://arxiv.org/abs/1309.3428} {arXiv:1309.3428
  [astro-ph.HE]} \BibitemShut {NoStop}%
\bibitem [{\citenamefont {Hooper}\ and\ \citenamefont
  {Slatyer}(2013)}]{hooperslatyer2013}%
  \BibitemOpen
  \bibfield  {author} {\bibinfo {author} {\bibfnamefont {Dan}\ \bibnamefont
  {Hooper}}\ and\ \bibinfo {author} {\bibfnamefont {Tracy~R.}\ \bibnamefont
  {Slatyer}},\ }\bibfield  {title} {\enquote {\bibinfo {title} {{Two Emission
  Mechanisms in the Fermi Bubbles: A Possible Signal of Annihilating Dark
  Matter}},}\ }\href@noop {} {\  (\bibinfo {year} {2013})},\ \Eprint
  {http://arxiv.org/abs/1302.6589} {arXiv:1302.6589 [astro-ph.HE]} \BibitemShut
  {NoStop}%
\bibitem [{\citenamefont {Huang}\ \emph {et~al.}(2013)\citenamefont {Huang},
  \citenamefont {Urbano},\ and\ \citenamefont {Xue}}]{Huang:2013}%
  \BibitemOpen
  \bibfield  {author} {\bibinfo {author} {\bibfnamefont {Wei-Chih}\
  \bibnamefont {Huang}}, \bibinfo {author} {\bibfnamefont {Alfredo}\
  \bibnamefont {Urbano}}, \ and\ \bibinfo {author} {\bibfnamefont {Wei}\
  \bibnamefont {Xue}},\ }\bibfield  {title} {\enquote {\bibinfo {title} {{Fermi
  Bubbles under Dark Matter Scrutiny. Part I: Astrophysical Analysis}},}\
  }\href@noop {} {\  (\bibinfo {year} {2013})},\ \Eprint
  {http://arxiv.org/abs/1307.6862} {arXiv:1307.6862 [hep-ph]} \BibitemShut
  {NoStop}%
\bibitem [{\citenamefont {Hooper}\ \emph {et~al.}(2013)\citenamefont {Hooper},
  \citenamefont {Cholis}, \citenamefont {Linden}, \citenamefont
  {Siegal-Gaskins},\ and\ \citenamefont {Slatyer}}]{Hooper:Pulsars}%
  \BibitemOpen
  \bibfield  {author} {\bibinfo {author} {\bibfnamefont {Dan}\ \bibnamefont
  {Hooper}}, \bibinfo {author} {\bibfnamefont {Ilias}\ \bibnamefont {Cholis}},
  \bibinfo {author} {\bibfnamefont {Tim}\ \bibnamefont {Linden}}, \bibinfo
  {author} {\bibfnamefont {Jennifer}\ \bibnamefont {Siegal-Gaskins}}, \ and\
  \bibinfo {author} {\bibfnamefont {Tracy}\ \bibnamefont {Slatyer}},\
  }\bibfield  {title} {\enquote {\bibinfo {title} {{Millisecond pulsars Cannot
  Account for the Inner Galaxy's GeV Excess}},}\ }\href {\doibase
  10.1103/PhysRevD.88.083009} {\bibfield  {journal} {\bibinfo  {journal}
  {Phys.Rev.}\ }\textbf {\bibinfo {volume} {D88}},\ \bibinfo {pages} {083009}
  (\bibinfo {year} {2013})},\ \Eprint {http://arxiv.org/abs/1305.0830}
  {arXiv:1305.0830 [astro-ph.HE]} \BibitemShut {NoStop}%
\bibitem [{\citenamefont {Linden}\ \emph {et~al.}(2012)\citenamefont {Linden},
  \citenamefont {Lovegrove},\ and\ \citenamefont {Profumo}}]{Linden:2012iv}%
  \BibitemOpen
  \bibfield  {author} {\bibinfo {author} {\bibfnamefont {Tim}\ \bibnamefont
  {Linden}}, \bibinfo {author} {\bibfnamefont {Elizabeth}\ \bibnamefont
  {Lovegrove}}, \ and\ \bibinfo {author} {\bibfnamefont {Stefano}\ \bibnamefont
  {Profumo}},\ }\bibfield  {title} {\enquote {\bibinfo {title} {{The Morphology
  of Hadronic Emission Models for the Gamma-Ray Source at the Galactic
  Center}},}\ }\href {\doibase 10.1088/0004-637X/753/1/41} {\bibfield
  {journal} {\bibinfo  {journal} {Astrophys.J.}\ }\textbf {\bibinfo {volume}
  {753}},\ \bibinfo {pages} {41} (\bibinfo {year} {2012})},\ \Eprint
  {http://arxiv.org/abs/1203.3539} {arXiv:1203.3539 [astro-ph.HE]} \BibitemShut
  {NoStop}%
\bibitem [{\citenamefont {Yusef-Zadeh}\ \emph {et~al.}(2013)\citenamefont
  {Yusef-Zadeh} \emph {et~al.}}]{yusef-zadehhewittwardle2013}%
  \BibitemOpen
  \bibfield  {author} {\bibinfo {author} {\bibfnamefont {F.}~\bibnamefont
  {Yusef-Zadeh}} \emph {et~al.},\ }\bibfield  {title} {\enquote {\bibinfo
  {title} {{Interacting Cosmic Rays with Molecular Clouds: A Bremsstrahlung
  Origin of Diffuse High Energy Emission from the Inner 2deg by 1deg of the
  Galactic Center}},}\ }\href {\doibase 10.1088/0004-637X/762/1/33} {\bibfield
  {journal} {\bibinfo  {journal} {Astrophys.J.}\ }\textbf {\bibinfo {volume}
  {762}},\ \bibinfo {pages} {33} (\bibinfo {year} {2013})},\ \Eprint
  {http://arxiv.org/abs/1206.6882} {arXiv:1206.6882 [astro-ph.HE]} \BibitemShut
  {NoStop}%
\bibitem [{\citenamefont {{Law}}\ \emph {et~al.}(2008)\citenamefont {{Law}},
  \citenamefont {{Yusef-Zadeh}}, \citenamefont {{Cotton}},\ and\ \citenamefont
  {{Maddalena}}}]{law2008}%
  \BibitemOpen
  \bibfield  {author} {\bibinfo {author} {\bibfnamefont {C.~J.}\ \bibnamefont
  {{Law}}}, \bibinfo {author} {\bibfnamefont {F.}~\bibnamefont
  {{Yusef-Zadeh}}}, \bibinfo {author} {\bibfnamefont {W.~D.}\ \bibnamefont
  {{Cotton}}}, \ and\ \bibinfo {author} {\bibfnamefont {R.~J.}\ \bibnamefont
  {{Maddalena}}},\ }\bibfield  {title} {\enquote {\bibinfo {title} {{Green Bank
  Telescope Multiwavelength Survey of the Galactic Center Region}},}\ }\href
  {\doibase 10.1086/533587} {\bibfield  {journal} {\bibinfo  {journal} {\apjs}\
  }\textbf {\bibinfo {volume} {177}},\ \bibinfo {pages} {255--274} (\bibinfo
  {year} {2008})},\ \Eprint {http://arxiv.org/abs/0801.4294} {arXiv:0801.4294}
  \BibitemShut {NoStop}%
\bibitem [{fer()}]{fermitools}%
  \BibitemOpen
  \href@noop {} {\enquote {\bibinfo {title} {Fermi science tools
  documentation},}\ }\bibinfo {howpublished}
  {\url{http://fermi.gsfc.nasa.gov/ssc/data/analysis/documentation/}}\BibitemShut
  {NoStop}%
\bibitem [{\citenamefont {Nolan}\ \emph {et~al.}(2012)\citenamefont {Nolan}
  \emph {et~al.}}]{2FGL}%
  \BibitemOpen
  \bibfield  {author} {\bibinfo {author} {\bibfnamefont {P.~L.}\ \bibnamefont
  {Nolan}} \emph {et~al.},\ }\bibfield  {title} {\enquote {\bibinfo {title}
  {{Fermi Large Area Telescope Second Source Catalog}},}\ }\href {\doibase
  10.1088/0067-0049/199/2/31} {\bibfield  {journal} {\bibinfo  {journal}
  {Astrophys.J.Suppl.}\ }\textbf {\bibinfo {volume} {199}},\ \bibinfo {pages}
  {31} (\bibinfo {year} {2012})},\ \Eprint {http://arxiv.org/abs/1108.1435}
  {arXiv:1108.1435 [astro-ph.HE]} \BibitemShut {NoStop}%
\bibitem [{\citenamefont {{Aharonian}}\ \emph {et~al.}(2006)\citenamefont
  {{Aharonian}} \emph {et~al.}}]{Aharonian:2006au}%
  \BibitemOpen
  \bibfield  {author} {\bibinfo {author} {\bibfnamefont {F.}~\bibnamefont
  {{Aharonian}}} \emph {et~al.},\ }\bibfield  {title} {\enquote {\bibinfo
  {title} {{Discovery of very-high-energy {$\gamma$}-rays from the Galactic
  Centre ridge}},}\ }\href {\doibase 10.1038/nature04467} {\bibfield  {journal}
  {\bibinfo  {journal} {\nat}\ }\textbf {\bibinfo {volume} {439}},\ \bibinfo
  {pages} {695--698} (\bibinfo {year} {2006})},\ \Eprint
  {http://arxiv.org/abs/arXiv:astro-ph/0603021} {arXiv:astro-ph/0603021}
  \BibitemShut {NoStop}%
\bibitem [{\citenamefont {Diffuse}\ and\ \citenamefont
  {Collaboration}(2009)}]{DGB}%
  \BibitemOpen
  \bibfield  {author} {\bibinfo {author} {\bibnamefont {Diffuse}}\ and\
  \bibinfo {author} {\bibfnamefont {Molecular Clouds Science Working Group
  Fermi Large Area~Telescope}\ \bibnamefont {Collaboration}},\ }\href@noop {}
  {\enquote {\bibinfo {title} {Description and caveats for the {LAT} team model
  of diffuse gamma-ray emission, version: gll\_iem\_v02.fit},}\ }\bibinfo
  {howpublished}
  {\url{http://fermi.gsfc.nasa.gov/ssc/data/access/lat/ring_for_FSSC_final4.pdf}}
  (\bibinfo {year} {2009})\BibitemShut {NoStop}%
\bibitem [{\citenamefont {Ackermann}\ \emph {et~al.}(2012)\citenamefont
  {Ackermann} \emph {et~al.}}]{ackermannajelloatwood2012}%
  \BibitemOpen
  \bibfield  {author} {\bibinfo {author} {\bibfnamefont {M.}~\bibnamefont
  {Ackermann}} \emph {et~al.},\ }\bibfield  {title} {\enquote {\bibinfo {title}
  {Fermi-lat observations of the diffuse gamma-ray emission: Implications for
  cosmic rays and the interstellar medium},}\ }\href
  {http://stacks.iop.org/0004-637X/750/i=1/a=3} {\bibfield  {journal} {\bibinfo
   {journal} {The Astrophysical Journal}\ }\textbf {\bibinfo {volume} {750}},\
  \bibinfo {pages} {3} (\bibinfo {year} {2012})}\BibitemShut {NoStop}%
\bibitem [{\citenamefont {Strong}\ and\ \citenamefont
  {Moskalenko}(1998)}]{GALPROP}%
  \BibitemOpen
  \bibfield  {author} {\bibinfo {author} {\bibfnamefont {A.W.}\ \bibnamefont
  {Strong}}\ and\ \bibinfo {author} {\bibfnamefont {I.V.}\ \bibnamefont
  {Moskalenko}},\ }\bibfield  {title} {\enquote {\bibinfo {title} {{Propagation
  of cosmic-ray nucleons in the galaxy}},}\ }\href {\doibase 10.1086/306470}
  {\bibfield  {journal} {\bibinfo  {journal} {Astrophys.J.}\ }\textbf {\bibinfo
  {volume} {509}},\ \bibinfo {pages} {212--228} (\bibinfo {year} {1998})},\
  \Eprint {http://arxiv.org/abs/astro-ph/9807150} {arXiv:astro-ph/9807150
  [astro-ph]} \BibitemShut {NoStop}%
\bibitem [{\citenamefont {Vladimirov}\ \emph {et~al.}(2011)\citenamefont
  {Vladimirov}, \citenamefont {Digel}, \citenamefont {Johannesson},
  \citenamefont {Michelson}, \citenamefont {Moskalenko} \emph
  {et~al.}}]{webrun}%
  \BibitemOpen
  \bibfield  {author} {\bibinfo {author} {\bibfnamefont {Andrey~E.}\
  \bibnamefont {Vladimirov}}, \bibinfo {author} {\bibfnamefont {Seth~W.}\
  \bibnamefont {Digel}}, \bibinfo {author} {\bibfnamefont {Gudlaugur}\
  \bibnamefont {Johannesson}}, \bibinfo {author} {\bibfnamefont {Peter~F.}\
  \bibnamefont {Michelson}}, \bibinfo {author} {\bibfnamefont {Igor~V.}\
  \bibnamefont {Moskalenko}},  \emph {et~al.},\ }\bibfield  {title} {\enquote
  {\bibinfo {title} {{GALPROP WebRun: an internet-based service for calculating
  galactic cosmic ray propagation and associated photon emissions}},}\ }\href
  {\doibase 10.1016/j.cpc.2011.01.017} {\bibfield  {journal} {\bibinfo
  {journal} {Comput.Phys.Commun.}\ }\textbf {\bibinfo {volume} {182}},\
  \bibinfo {pages} {1156--1161} (\bibinfo {year} {2011})},\ \Eprint
  {http://arxiv.org/abs/1008.3642} {arXiv:1008.3642 [astro-ph.HE]} \BibitemShut
  {NoStop}%
\bibitem [{\citenamefont {Jeltema}\ and\ \citenamefont
  {Profumo}(2008)}]{Profumo1}%
  \BibitemOpen
  \bibfield  {author} {\bibinfo {author} {\bibfnamefont {Tesla~E.}\
  \bibnamefont {Jeltema}}\ and\ \bibinfo {author} {\bibfnamefont {Stefano}\
  \bibnamefont {Profumo}},\ }\bibfield  {title} {\enquote {\bibinfo {title}
  {{Fitting the Gamma-Ray Spectrum from Dark Matter with DMFIT: GLAST and the
  Galactic Center Region}},}\ }\href {\doibase 10.1088/1475-7516/2008/11/003}
  {\bibfield  {journal} {\bibinfo  {journal} {JCAP}\ }\textbf {\bibinfo
  {volume} {0811}},\ \bibinfo {pages} {003} (\bibinfo {year} {2008})},\ \Eprint
  {http://arxiv.org/abs/0808.2641} {arXiv:0808.2641 [astro-ph]} \BibitemShut
  {NoStop}%
\bibitem [{\citenamefont {Abdo}\ \emph {et~al.}(2013)\citenamefont {Abdo} \emph
  {et~al.}}]{FermiPulsarsCatalog}%
  \BibitemOpen
  \bibfield  {author} {\bibinfo {author} {\bibfnamefont {A.A.}\ \bibnamefont
  {Abdo}} \emph {et~al.} (\bibinfo {collaboration} {The Fermi-LAT
  collaboration}),\ }\bibfield  {title} {\enquote {\bibinfo {title} {{The
  Second Fermi Large Area Telescope Catalog of Gamma-ray Pulsars}},}\ }\href
  {\doibase 10.1088/0067-0049/208/2/17} {\bibfield  {journal} {\bibinfo
  {journal} {Astrophys.J.Suppl.}\ }\textbf {\bibinfo {volume} {208}},\ \bibinfo
  {pages} {17} (\bibinfo {year} {2013})},\ \Eprint
  {http://arxiv.org/abs/1305.4385} {arXiv:1305.4385 [astro-ph.HE]} \BibitemShut
  {NoStop}%
\bibitem [{\citenamefont {{Abdo}}\ \emph {et~al.}(2010)\citenamefont {{Abdo}}
  \emph {et~al.}}]{FermiGlobularClusters}%
  \BibitemOpen
  \bibfield  {author} {\bibinfo {author} {\bibfnamefont {A.~A.}\ \bibnamefont
  {{Abdo}}} \emph {et~al.},\ }\bibfield  {title} {\enquote {\bibinfo {title}
  {{A population of gamma-ray emitting globular clusters seen with the Fermi
  Large Area Telescope}},}\ }\href {\doibase 10.1051/0004-6361/201014458}
  {\bibfield  {journal} {\bibinfo  {journal} {\aap}\ }\textbf {\bibinfo
  {volume} {524}},\ \bibinfo {eid} {A75} (\bibinfo {year} {2010})},\ \Eprint
  {http://arxiv.org/abs/1003.3588} {arXiv:1003.3588 [astro-ph.GA]} \BibitemShut
  {NoStop}%
\bibitem [{\citenamefont {{The Fermi-LAT
  Collaboration}}(2013)}]{FermiDwarf4yr}%
  \BibitemOpen
  \bibfield  {author} {\bibinfo {author} {\bibnamefont {{The Fermi-LAT
  Collaboration}}},\ }\bibfield  {title} {\enquote {\bibinfo {title} {{Dark
  Matter Constraints from Observations of 25 Milky Way Satellite Galaxies with
  the Fermi Large Area Telescope}},}\ }\href@noop {} {\bibfield  {journal}
  {\bibinfo  {journal} {ArXiv e-prints}\ } (\bibinfo {year} {2013})},\ \Eprint
  {http://arxiv.org/abs/1310.0828} {arXiv:1310.0828 [astro-ph.HE]} \BibitemShut
  {NoStop}%
\bibitem [{\citenamefont {{Ackermann}}\ \emph {et~al.}(2012)\citenamefont
  {{Ackermann}} \emph {et~al.}}]{FermiInnerGalaxy2012}%
  \BibitemOpen
  \bibfield  {author} {\bibinfo {author} {\bibfnamefont {M.}~\bibnamefont
  {{Ackermann}}} \emph {et~al.},\ }\bibfield  {title} {\enquote {\bibinfo
  {title} {{Constraints on the Galactic Halo Dark Matter from Fermi-LAT Diffuse
  Measurements}},}\ }\href {\doibase 10.1088/0004-637X/761/2/91} {\bibfield
  {journal} {\bibinfo  {journal} {\apj}\ }\textbf {\bibinfo {volume} {761}},\
  \bibinfo {eid} {91} (\bibinfo {year} {2012})},\ \Eprint
  {http://arxiv.org/abs/1205.6474} {arXiv:1205.6474 [astro-ph.CO]} \BibitemShut
  {NoStop}%
\bibitem [{\citenamefont {Wilks}(1938)}]{wilks}%
  \BibitemOpen
  \bibfield  {author} {\bibinfo {author} {\bibfnamefont {S.~S.}\ \bibnamefont
  {Wilks}},\ }\bibfield  {title} {\enquote {\bibinfo {title} {{The Large-Sample
  Distribution of the Likelihood Ratio for Testing Composite Hypotheses}},}\
  }\href {\doibase 10.1214/aoms/1177732360} {\bibfield  {journal} {\bibinfo
  {journal} {Ann.Math.Stat.}\ }\textbf {\bibinfo {volume} {9}},\ \bibinfo
  {pages} {60} (\bibinfo {year} {1938})}\BibitemShut {NoStop}%
\bibitem [{\citenamefont {Mattox}\ \emph {et~al.}(1996)\citenamefont {Mattox},
  \citenamefont {Bertsch}, \citenamefont {Chiang}, \citenamefont {Dingus},
  \citenamefont {Digel} \emph {et~al.}}]{mattox}%
  \BibitemOpen
  \bibfield  {author} {\bibinfo {author} {\bibfnamefont {J.R.}\ \bibnamefont
  {Mattox}}, \bibinfo {author} {\bibfnamefont {D.L.}\ \bibnamefont {Bertsch}},
  \bibinfo {author} {\bibfnamefont {J.}~\bibnamefont {Chiang}}, \bibinfo
  {author} {\bibfnamefont {B.L.}\ \bibnamefont {Dingus}}, \bibinfo {author}
  {\bibfnamefont {S.W.}\ \bibnamefont {Digel}},  \emph {et~al.},\ }\bibfield
  {title} {\enquote {\bibinfo {title} {{The Likelihood Analysis of EGRET
  Data}},}\ }\href@noop {} {\bibfield  {journal} {\bibinfo  {journal}
  {Astrophys.J.}\ }\textbf {\bibinfo {volume} {461}},\ \bibinfo {pages} {396}
  (\bibinfo {year} {1996})}\BibitemShut {NoStop}%
\bibitem [{\citenamefont {Uchiyama}\ \emph {et~al.}(2012)\citenamefont
  {Uchiyama}, \citenamefont {Funk}, \citenamefont {Katagiri}, \citenamefont
  {Katsuta}, \citenamefont {Lemoine-Goumard} \emph {et~al.}}]{Uchiyama&Funk}%
  \BibitemOpen
  \bibfield  {author} {\bibinfo {author} {\bibfnamefont {Yasunobu}\
  \bibnamefont {Uchiyama}}, \bibinfo {author} {\bibfnamefont {Stefan}\
  \bibnamefont {Funk}}, \bibinfo {author} {\bibfnamefont {Hideaki}\
  \bibnamefont {Katagiri}}, \bibinfo {author} {\bibfnamefont {Junichiro}\
  \bibnamefont {Katsuta}}, \bibinfo {author} {\bibfnamefont {Marianne}\
  \bibnamefont {Lemoine-Goumard}},  \emph {et~al.},\ }\bibfield  {title}
  {\enquote {\bibinfo {title} {{Fermi-LAT Discovery of GeV Gamma-ray Emission
  from the Vicinity of SNR W44}},}\ }\href@noop {} {\  (\bibinfo {year}
  {2012})},\ \Eprint {http://arxiv.org/abs/1203.3234} {arXiv:1203.3234
  [astro-ph.HE]} \BibitemShut {NoStop}%
\bibitem [{\citenamefont {Ackermann}\ \emph {et~al.}(2013)\citenamefont
  {Ackermann} \emph {et~al.}}]{Ackermann:2013wqa}%
  \BibitemOpen
  \bibfield  {author} {\bibinfo {author} {\bibfnamefont {M.}~\bibnamefont
  {Ackermann}} \emph {et~al.} (\bibinfo {collaboration} {Fermi-LAT
  Collaboration}),\ }\bibfield  {title} {\enquote {\bibinfo {title} {{Detection
  of the Characteristic Pion-Decay Signature in Supernova Remnants}},}\ }\href
  {\doibase 10.1126/science.1231160} {\bibfield  {journal} {\bibinfo  {journal}
  {Science}\ }\textbf {\bibinfo {volume} {339}},\ \bibinfo {pages} {807}
  (\bibinfo {year} {2013})},\ \Eprint {http://arxiv.org/abs/1302.3307}
  {arXiv:1302.3307 [astro-ph.HE]} \BibitemShut {NoStop}%
\bibitem [{\citenamefont {James}\ and\ \citenamefont {Roos}(1975)}]{minuit}%
  \BibitemOpen
  \bibfield  {author} {\bibinfo {author} {\bibfnamefont {F.}~\bibnamefont
  {James}}\ and\ \bibinfo {author} {\bibfnamefont {M.}~\bibnamefont {Roos}},\
  }\bibfield  {title} {\enquote {\bibinfo {title} {{Minuit: A System for
  Function Minimization and Analysis of the Parameter Errors and
  Correlations}},}\ }\href {\doibase 10.1016/0010-4655(75)90039-9} {\bibfield
  {journal} {\bibinfo  {journal} {Comput.Phys.Commun.}\ }\textbf {\bibinfo
  {volume} {10}},\ \bibinfo {pages} {343--367} (\bibinfo {year}
  {1975})}\BibitemShut {NoStop}%
\bibitem [{\citenamefont {Jones}\ \emph {et~al.}(2001)\citenamefont {Jones},
  \citenamefont {Oliphant}, \citenamefont {Peterson} \emph {et~al.}}]{scipy}%
  \BibitemOpen
  \bibfield  {author} {\bibinfo {author} {\bibfnamefont {Eric}\ \bibnamefont
  {Jones}}, \bibinfo {author} {\bibfnamefont {Travis}\ \bibnamefont
  {Oliphant}}, \bibinfo {author} {\bibfnamefont {Pearu}\ \bibnamefont
  {Peterson}},  \emph {et~al.},\ }\href {http://www.scipy.org/} {\enquote
  {\bibinfo {title} {{SciPy}: Open source scientific tools for {Python}},}\ }
  (\bibinfo {year} {2001})\BibitemShut {NoStop}%
\bibitem [{\citenamefont {{Schlickeiser}}(2002)}]{Schlickeiser}%
  \BibitemOpen
  \bibfield  {author} {\bibinfo {author} {\bibfnamefont {R.}~\bibnamefont
  {{Schlickeiser}}},\ }\href@noop {} {\emph {\bibinfo {title} {Cosmic ray
  astrophysics / Reinhard Schlickeiser, Astronomy and Astrophysics Library;
  Physics and Astronomy Online Library.~Berlin: Springer.~ISBN 3-540-66465-3,
  2002, XV + 519 pp.}}}\ (\bibinfo {year} {2002})\BibitemShut {NoStop}%
\bibitem [{\citenamefont {Bethe}\ and\ \citenamefont {Heitler}(1934)}]{Bethe}%
  \BibitemOpen
  \bibfield  {author} {\bibinfo {author} {\bibfnamefont {H.}~\bibnamefont
  {Bethe}}\ and\ \bibinfo {author} {\bibfnamefont {W.}~\bibnamefont
  {Heitler}},\ }\bibfield  {title} {\enquote {\bibinfo {title} {{On the
  Stopping of fast particles and on the creation of positive electrons}},}\
  }\href {\doibase 10.1098/rspa.1934.0140} {\bibfield  {journal} {\bibinfo
  {journal} {Proc.Roy.Soc.Lond.}\ }\textbf {\bibinfo {volume} {A146}},\
  \bibinfo {pages} {83--112} (\bibinfo {year} {1934})}\BibitemShut {NoStop}%
\bibitem [{\citenamefont {{Schlickeiser}}\ and\ \citenamefont
  {{Thielheim}}(1978)}]{SchlickeiserThielheim1978}%
  \BibitemOpen
  \bibfield  {author} {\bibinfo {author} {\bibfnamefont {R.}~\bibnamefont
  {{Schlickeiser}}}\ and\ \bibinfo {author} {\bibfnamefont {K.~O.}\
  \bibnamefont {{Thielheim}}},\ }\bibfield  {title} {\enquote {\bibinfo {title}
  {{Non-thermal electron bremsstrahlung in the Galactic disk}},}\ }\href@noop
  {} {\bibfield  {journal} {\bibinfo  {journal} {\mnras}\ }\textbf {\bibinfo
  {volume} {182}},\ \bibinfo {pages} {103--109} (\bibinfo {year}
  {1978})}\BibitemShut {NoStop}%
\bibitem [{\citenamefont {Beringer}\ \emph {et~al.}(2012)\citenamefont
  {Beringer} \emph {et~al.}}]{pdg}%
  \BibitemOpen
  \bibfield  {author} {\bibinfo {author} {\bibfnamefont {J.}~\bibnamefont
  {Beringer}} \emph {et~al.} (\bibinfo {collaboration} {Particle Data Group}),\
  }\bibfield  {title} {\enquote {\bibinfo {title} {{Review of Particle Physics
  (RPP)}},}\ }\href {\doibase 10.1103/PhysRevD.86.010001} {\bibfield  {journal}
  {\bibinfo  {journal} {Phys.Rev.}\ }\textbf {\bibinfo {volume} {D86}},\
  \bibinfo {pages} {010001} (\bibinfo {year} {2012})}\BibitemShut {NoStop}%
\bibitem [{\citenamefont {Rybicki}\ and\ \citenamefont
  {Lightman}(1985)}]{Rybicki}%
  \BibitemOpen
  \bibfield  {author} {\bibinfo {author} {\bibfnamefont {George~B.}\
  \bibnamefont {Rybicki}}\ and\ \bibinfo {author} {\bibfnamefont {Alan~P.}\
  \bibnamefont {Lightman}},\ }\bibfield  {title} {\enquote {\bibinfo {title}
  {{Radiative Processes in Astrophysics}},}\ }\href@noop {} {\  (\bibinfo
  {year} {1985})}\BibitemShut {NoStop}%
\end{thebibliography}%

\appendix

\section{Nonthermal Bremsstrahlung Spectrum  }
\label{sec:bremsstrahlungFormulae}
In this appendix we discuss the  relevant synchrotron and bremsstrahlung formula which were used in the analysis of the current article and Ref.~\citep{yusef-zadehhewittwardle2013}. As these formulas are only briefly alluded to  in Ref.~\citep{yusef-zadehhewittwardle2013} we provide more details here which are based on notes kindly supplied to us by Prof.~Mark Wardle. 

Relativistic cosmic ray electrons that are deflected in the Coulomb field of nuclei in molecular clouds at the Galactic Center emit bremsstrahlung $\gamma$-ray photons~\citep{Schlickeiser}. In this region the ionized gas component contribution to the radiation process can be neglected~\citep{Schlickeiser}. By considering this the differential cross section for the bremsstrahlung interaction~\citep{Bethe,SchlickeiserThielheim1978} can be written 
as\footnote{We note that Eq.~4.4.1 of Ref.~\citep{Schlickeiser} is missing a factor of $1/E_\gamma$. }
\begin{eqnarray}\nonumber
\sigma(E_{\gamma},\gamma)&=&
\frac{3}{8\pi E_{\gamma}}\; \alpha\; \sigma_{T}\; \Phi_{\rm H}\left[\frac{4}{3}-\frac{4}{3}\frac{E_{\gamma}}{E} +\left(\frac{E_{\gamma}}{E}\right)^2 \right] \nonumber \\
&& {\rm cm}^2\  {\rm eV}^{-1}\ 
\label{eq:cross section}
\end{eqnarray}   
where $E_{\gamma}$ is the photon energy, $E=\gamma mc^2$ the relativistic electron energy, $\alpha=1/137.0$ the fine structure constant, $\sigma_{T}=6.652\times 10^{-25}$ cm$^2$ the Thomson cross section and $\Phi_{\rm H}\simeq 45$ the scattering function assumed to be in the strong-shielding limit which is appropriate for relativistic electrons. We take the invariant electron mass to be $m=9.109\times10^{-28}$ g~\citep{pdg} and for the speed of light we use $c=2.998\times 10^{10}$ cm/s.

The nonthermal electron bremsstrahlung omnidirectional source function produced by a single relativistic electron in a medium dominated by atomic and molecular hydrogen nuclei of corresponding number density $n_{\rm H}=n_{\text{HI}}+2n_{\text{H$_2$}}$ is given by~\citep{Schlickeiser}

\begin{eqnarray} \nonumber
q(E_{\gamma})&=&c\; n_{\rm H}\int^{+\infty}_{E_{\rm L}}d\gamma \; n(\gamma)\; \sigma(E_{\gamma},\gamma)\\
&& {\rm photons}\ {\rm cm}^{-3}\ {\rm s}^{-1}\ {\rm eV}^{-1}
\label{eq:defsourcefunction}
\end{eqnarray}
where $E_{\rm L}=\text{max}(E_{\gamma},E_1)$ and $E_1=\gamma_1mc^2$ represents a low-energy cutoff in the electron distribution. The energy distribution function $n(\gamma)$ of the radiating relativistic electrons is assumed to follow a broken power law of the form

\begin{eqnarray}
n(\gamma)=
  \begin{cases}
    n_1\;\gamma^{-p_1} &\text{if $1\leq \gamma \leq \gamma_b$} \\
n_2\; \gamma^{-p_2} &\text{if $ \gamma \geq \gamma_b$},
  \end{cases}
\quad {\rm cm}^{-3}
\label{eq:ngamma}  
\end{eqnarray}
with $n_1\gamma_b^{-p_1}=n_2\gamma_b^{-p_2}\equiv n_b$ and $E_b=\gamma_b mc^2$ the break energy. After substituting Eq.~\eqref{eq:cross section} and~\eqref{eq:ngamma} into Eq.~\eqref{eq:defsourcefunction} and solving the corresponding integrals we get for the omnidirectional source function

\begin{eqnarray}\nonumber
q(E_{\gamma})&=&\frac{3\;\alpha\; \sigma_T}{8\pi }\Phi_{\rm H}\;n_{\rm H}\frac{n_b}{mc}\;J\left(E_{\gamma}/E_b\right) \\
& & \text{photons  cm$^{-3}$ s$^{-1}$eV$^{-1}$,} 
\label{eq:explicitsourcefunction}
\end{eqnarray}
where 
\begin{widetext} 
\begin{eqnarray}
J(x)= 
  \begin{cases}
    \left[I_{p_1}(1)-I_{p_1}(1/x)\right]x^{-p_1} + I_{p_2}(1/x)\;x^{-p_2} & \text{for $x < 1$} \\
I_{p_2}(1)\;x^{-p_2} & \text{for $x \geq 1$},
  \end{cases} 
\label{eq:Jfunction}  
\end{eqnarray}
\end{widetext}
and 

\begin{eqnarray}\nonumber
I_{p_i}(x)&=&\frac{1}{3}x^{-p_i}\left(\frac{3}{x+p_i\;x}+\frac{4x}{p_i-1}-\frac{4}{p_i} \right)\\&& \text{for $i=1,2$.} 
\label{eq:Ifunction}
\end{eqnarray}

In Ref.~\citep{yusef-zadehhewittwardle2013} it was argued that the morphological distribution of diffuse radiation from the GC measured at 1.45 GHz, GeV and TeV energies is correlated. Importantly, in that work it was assumed that all the synchrotron emitting electrons are interacting with the molecular clouds. It is thus interesting to find an expression for the bremsstrahlung spectrum in terms of the synchrotron flux.
      
Synchrotron emission at frequency $\nu$ (taken to be the characteristic synchrotron radiation frequency) is associated with particles of energy (see Eq.~4.1.9 of \citep{Schlickeiser})

\begin{eqnarray}\nonumber
E_{\nu}&=&\gamma_{\nu}mc^2\\
       &=&\left(\frac{4\pi mc\nu}{3eB} \right)^{1/2}mc^2 \nonumber \\
       &=&7.89\left(\frac{B}{\rm \mu G}\right)^{-1/2}\left(\frac{\nu}{\rm GHz}\right)^{1/2} \quad {\rm GeV}\, .
\label{eq:Enu}
\end{eqnarray}
where the electron charge is $e=4.803\times 10^{-10}$ statcoulomb.
 Eq.~\ref{eq:Enu} can be rewritten as
\begin{eqnarray}\nonumber
\gamma_{\nu}=\left(\frac{\nu}{\nu_B}\right)^{1/2} \text{where}\   \nu_B=\frac{3eB}{4\pi mc}=4.20\left( \frac{B}{\rm \mu G}\right)\quad {\rm Hz}.\\
\end{eqnarray}
We therefore obtain that particles radiating at the synchrotron break frequency $\nu_b$ obey the relation
\begin{eqnarray}
\nu_b=\nu_B\gamma^2_b\quad {\rm Hz}.
\end{eqnarray}

The synchrotron emission coefficient resulting from an electron spectrum that is a simple power law can be obtained from Eq.(6.36) of Ref.~\citep{Rybicki} which can be rewritten as
\begin{eqnarray} \nonumber
j_{\nu}=\frac{1}{\sqrt{3}}f_j(p)\frac{e^2}{c}\;\nu_B\; \gamma_{\nu}\;n\left(\gamma_{\nu}\right)
\quad {\rm erg}\ {\rm cm}^{-3}\, {\rm ster}^{-1}\ {\rm s}^{-1}\ {\rm Hz}^{-1} \\
\label{eq:Jnu}
\end{eqnarray}

where 

\begin{eqnarray}
f_j(p)=\frac{2^{(p-1)/2}}{p+1}\Gamma\left(\frac{3p-1}{12}\right)\Gamma\left(\frac{3p+19}{12}\right),
\end{eqnarray}
and $\Gamma(z)$ is the Gamma function. Using Eq.~\ref{eq:Jnu} we can estimate the emission coefficient  for a broken power law at the synchrotron break frequency 

\begin{eqnarray}
j_b=\tilde{j}n_b\quad {\rm erg}\ {\rm cm}^{-3}\ {\rm ster}^{-1}\ {\rm s}^{-1}\ {\rm Hz}^{-1},
\end{eqnarray} 
where
\begin{eqnarray}
\tilde{j}=\frac{1}{\sqrt{3}}\bar{f}_j(p)\frac{e^2}{c}\sqrt{\nu_B\nu_b}\quad  {\rm erg}\ {\rm ster}^{-1}\ {\rm s}^{-1}\ {\rm Hz}^{-1},
\end{eqnarray} 
and 
\begin{eqnarray}
\bar{f}_j(p)=\frac{1}{2}\left[f_j(p_1)+f_j(p_2)\right].
\end{eqnarray}
where $p_1$ and $p_2$ are the broken power law spectral indices before and after the break.    

Then the spectral value of synchrotron radiation from a source of volume $V$ at a distance $d$ at the break frequency is given by
\begin{eqnarray}
S_b=4\pi \frac{ \tilde{j} n_b V}{4 \pi d^2}\quad {\rm Jy}.
\label{eq:Sb}
\end{eqnarray}

Finally, the source function given in Eq.~\ref{eq:explicitsourcefunction} is multiplied by a factor $V$ to get the photon luminosity spectrum and then divided by  $4\pi d^2$ to obtain the bremsstrahlung photon flux spectrum and using the results of Eq~\ref{eq:Sb} we thus get for the bremsstrahlung $\gamma$-ray spectrum

\begin{eqnarray}\nonumber
\frac{dN_{\text{brem}}}{dE_{\gamma}}&=&\frac{3\;\alpha\; \sigma_T}{32\pi^2 \tilde{j} }\Phi_{\rm H}\;n_{\rm H}\frac{S_b}{mc}\;J\left(E_{\gamma}/E_b\right) \\
& & \text{photons erg$^{-1}$ cm$^{-2}$ s$^{-1}$,} 
\label{eq:bremsstrahlungSpectra}
\end{eqnarray}
where $J\left(E_{\gamma}/E_b\right)$ is given by Eq.~\ref{eq:Jfunction}.

\begin{table*}[h!]
\section{Gamma-Ray Excess Data }
\label{sec:data2}

\begin{ruledtabular}
\begin{tabular}{|c|c|c|c|c|c|}
\centering 
$E_{\rm min}$ [GeV] &    $E_{\rm max}$ [GeV] &    $dN/dE$ [GeV$^{-1}$ cm$^{-2}$ s$^{-1}$]  &    Stat.\ Error [GeV$^{-1}$ cm$^{-2}$ s$^{-1}$] &    Syst.\ Error [GeV$^{-1}$ cm$^{-2}$ s$^{-1}$] & TS \\ \hline
0.30&    0.40 &    $1.11\times 10^{-6}$ &    $6.41\times 10^{-7}$ &    $3.92\times 10^{-7}$ &3.06 \\
0.40&    0.54 &    $7.17\times 10^{-7}$ &    $6.71\times 10^{-8}$ &    $2.84\times 10^{-7}$ &5.93 \\
0.54&    0.72 &    $4.28\times 10^{-7}$ &    $4.54\times 10^{-8}$ &    $1.26\times 10^{-7}$ &10.45\\
0.72&    0.97 &    $3.02\times 10^{-7}$ &    $3.05\times 10^{-8}$ &    $7.98\times 10^{-8}$ &25.21\\
0.97&    1.29 &    $1.95\times 10^{-7}$ &    $2.57\times 10^{-8}$ &    $3.80\times 10^{-8}$ &49.1\\
1.29&    1.73 &    $1.23\times 10^{-7}$ &    $1.81\times 10^{-8}$ &    $1.83\times 10^{-8}$ &73.51\\
1.73&    2.32 &    $5.44\times 10^{-8}$ &    $1.48\times 10^{-8}$ &    $9.38\times 10^{-9}$ &63.94\\
2.32&    3.11 &    $3.39\times 10^{-8}$ &    $7.25\times 10^{-9}$ &    $4.35\times 10^{-9}$ &84.03\\
3.11&    4.16 &    $1.33\times 10^{-8}$ &    $6.11\times 10^{-9}$ &    $1.98\times 10^{-9}$ &43.3\\
4.16&    5.57 &    $3.42\times 10^{-9}$ &    $3.17\times 10^{-9}$ &    $3.17\times 10^{-9}$ &10.46\\
5.57&    7.47 &    $2.26\times 10^{-9}$ &    $6.97\times 10^{-10}$ &    $3.01\times 10^{-10}$ & 14.14\\
7.47&    10.00 &    $1.21\times 10^{-9}$ &    $2.51\times 10^{-10}$ &    $2.40\times 10^{-10}$ &12.82 \\
10.00&    100.00 &    $-$ &    $-$ &    $-$ & 0.12 

\end{tabular}
\end{ruledtabular}

\caption{\label{tab:DMdata20-cmMap} NFW$_{1.2}^2$ spectrum and corresponding statistical and systematic errors using model 1 in Table~\ref{tab:LogLikelihoods}. The spectral points $dN/dE$ were obtained at the logarithmic midpoint of each band.
 }
\end{table*}

\begin{table*}[h!]
\begin{ruledtabular}
\begin{tabular}{|c|c|c|c|c|c|}
\centering 
$E_{\rm min}$ [GeV] &    $E_{\rm max}$ [GeV] &    $dN/dE$ [GeV$^{-1}$ cm$^{-2}$ s$^{-1}$]  &    Stat. Error [GeV$^{-1}$ cm$^{-2}$ s$^{-1}$] &    Syst. Error [GeV$^{-1}$ cm$^{-2}$ s$^{-1}$] & TS \\ \hline
0.30&    0.40 &    $-$ &    $-$ &    $-$ &0.0 \\
0.40&    0.54 &    $2.70\times 10^{-7}$ &    $7.19\times 10^{-8}$ &    $9.21\times 10^{-8}$ &5.04 \\
0.54&    0.72 &    $1.24\times 10^{-7}$ &    $3.41\times 10^{-8}$ &    $5.57\times 10^{-8}$ &5.42\\
0.72&    0.97 &    $1.14\times 10^{-7}$ &    $2.13\times 10^{-8}$ &    $4.83\times 10^{-8}$ &20.71\\
0.97&    1.29 &    $5.32\times 10^{-8}$ &    $1.11\times 10^{-8}$ &    $4.35\times 10^{-8}$ &18.53\\
1.29&    1.73 &    $2.45\times 10^{-8}$ &    $7.20\times 10^{-9}$ &    $9.16\times 10^{-9}$ &16.49\\
1.73&    2.32 &    $1.39\times 10^{-8}$ &    $3.19\times 10^{-9}$ &    $7.09\times 10^{-9}$ &19.42\\
2.32&    3.11 &    $7.63\times 10^{-9}$ &    $1.71\times 10^{-9}$ &    $1.98\times 10^{-9}$ &22.89\\
3.11&    4.16 &    $5.04\times 10^{-9}$ &    $1.39\times 10^{-9}$ &    $8.02\times 10^{-10}$ &33.45\\
4.16&    5.57 &    $2.05\times 10^{-9}$ &    $3.83\times 10^{-10}$ &    $3.68\times 10^{-10}$ &21.57\\
5.57&    7.47 &    $7.67\times 10^{-10}$ &    $2.64\times 10^{-10}$ &    $1.59\times 10^{-10}$ & 8.82\\
7.47&    10.00 &    $2.37\times 10^{-10}$ &    $1.10\times 10^{-10}$ &    $8.06\times 10^{-11}$ &3.02 \\
10.00&    100.00 &   $3.53\times 10^{-12}$ &    $1.90\times 10^{-11}$ &    $5.69\times 10^{-12}$ & 1.06

\end{tabular}
\end{ruledtabular}

\caption{\label{tab:Ridgedata20-cmMapNOArcandSgrB} 
Galactic Ridge spectrum generated using model 2 in Table~\ref{tab:LogLikelihoods}.
 }
\end{table*}

\end{document}